%% file: MAIN.tex
\documentclass[11pt]{article}
\usepackage{preamble}
\usepackage{hyperref}
\usepackage{csquotes}
\usepackage{nameref}
\usepackage{tiffany}
\newcommand{\recommendation}[2]{\noindent\fbox{\parbox{\textwidth}{  \paragraph{Recommendation #1: } #2 \\} } \\ \vspace{0.2in}}

\title{Climate of the Field: Snowmass 2021}

\author[1]{Erin V Hansen\orcidlink{ 0000-0002-2271-8078}}
\affil[1]{Department of Physics, University of California, Berkeley, CA}

\author[2]{Erica Smith\orcidlink{0000-0002-2467-8364}}
\affil[2]{Indiana University, Bloomington, IN}

\author[3]{Deborah Bard\orcidlink{0000-0002-5162-5153}}
\affil[3]{National Energy Research Scientific Computing Center (NERSC), LBNL, CA}
\author[4]{Matthew Bellis\orcidlink{0000-0002-6353-6043}}
\affil[4]{Siena College, Loudonville, NY}

\author[5]{Jessica Esquivel}
\affil[5]{Fermi National Accelerator Laboratory, Batavia, IL}

\author[6,7]{Tiffany R. Lewis\orcidlink{0000-0002-9854-1432}}
\affil[6]{Astroparticle Physics Laboratory, NASA Goddard Space Flight Center, Greenbelt, MD}
\affil[7]{NASA Postdoctoral Program Fellow}

\author[8]{Cameron Geddes\orcidlink{0000-0002-3763-9743}}
\affil[8]{Lawrence Berkeley National Laboratory, Berkeley CA}
\author[5]{Cindy Joe}

\author[8]{Alex G. Kim\orcidlink{0000-0001-6315-8743}}
\author[8]{Asmita Patel}

\author[5]{Vitaly Pronskikh\orcidlink{0000-0002-5181-7497}}

\date{Most recent update: \today}

\begin{document}
\maketitle

\clearpage

\begin{abstract} 

How are formal policies put in place to create an inclusive, equitable, safe environment? How do these differ between different communities of practice (institutions, labs, collaborations, working groups)? What policies towards a more equitable community are working? For those that aren’t working, what external support is needed in order to make them more effective?

We present a discussion of the current climate of the field in high energy particle physics and astrophysics (HEPA), as well as current efforts toward making the community a more diverse, inclusive, and equitable environment. We also present issues facing both institutions and HEPA collaborations, with a set of interviews with a selection of HEPA collaboration DEI leaders.

We encourage the HEPA community and the institutions \& agencies that support it to think critically about the prioritization of \textit{people} in HEPA over the coming decade, and what resources and policies need to be in place in order to protect and elevate minoritized populations within the HEPA community. 

\end{abstract}

\newpage
\begin{center}
    {\Large Summary of Recommendations from this Report}
\end{center}
\input{SummaryOfRecs}

\newpage
\tableofcontents 

\clearpage
\ \ 
\vspace{3in}
\begin{displayquote}
    ``I’ll leave you with this take-home: `Diverse perspectives yield the best science' is a true statement, but it’s one that commodifies the lived experience of marginalized people by reducing them to their contributions to productivity. It’s a capitalistic framework that shirks the basic truth that cultivating a field where the norm is respecting the humanity and validity of all people is the right thing to do for no reason other than that it is right. If this is not enough of a justification for you, you are the problem."
    
    \hfill \#BlackInAstro Experiences: KeShawn Ivory \footnote{https://astrobites.org/2020/06/19/black-in-astro-keshawn-ivory/}
\end{displayquote}

\clearpage
\section{Introduction: The Climate of HEPA Needs Work}\label{sec:Introduction}
\input{introduction}

\clearpage
\section{Context: Reports from other communities} \label{sec:OtherReports}
\input{OtherClimateReports}


\clearpage
\section{Institution Policies in Practice: Designed to serve institutions, not people}
\label{sec:Institutions}
\input{InstitutionPolicies}


\clearpage
\section{Collaboration Systems: The best of efforts are not mitigating harm} \label{sec:Collaborations}
\input{CollaborationPolicies}




\newpage
\section{Scientists \& non-Scientists: Participatory Injustice} \label{sec:EngineersAndScientists}
\input{EngineeringRoles}

\newpage
\section{Equity in information sharing / HEPA software}\label{sec:InfoSharingHEPSoftware}
\input{HEPsoftware}

\newpage
\section{Throughlines and Additional Topics}\label{sec:NonCodeTopics}

\input{NonCodeTopics}

\newpage
\section{Collaboration Services}\label{sec:CollaborationServices}
\input{CollaborationServices}


\newpage
\section{Concluding Remarks}\label{sec:Conclusion}
We hope that this work has brought up new questions within HEPA communities and the institutions and agencies that support them --- new questions about old topics. We acknowledge that the themes that have arisen as part of this work are not novel, and yet we still must speak on them. As such, we will reiterate our full-throated support of the recommendations provided by Nord \textit{et al.} in the ``Culture change is necessary, and it requires
strategic planning" letter of intent submitted as part of the Snowmass planning process\cite{LOI_StrategicPlanning}; these recommendations are as follows: 

\vspace{.3in}
\recommendation{1.0}{HEPA communities must employ the use of robust strategic planning procedures, including a full re-envisioning of science workplace norms and culture.}

\recommendation{2.0}{HEPA communities must implement new modes of community organizing and decision-making that promote agency and leadership from all stakeholders within the scientific community.}

\recommendation{3.0}{HEPA communities must engage in partnership with scholars, professionals, and other experts in several disciplines, including but not limited to anti-racism, critical race theory, and social science.}

We encourage HEPA communities and the institutions and agencies that support them to reflect on what is necessary to reinvent the physics workplace, and then \textit{act} with policy, with funding, and with person-power toward these ends. Our community has a chance to, and realistically must go beyond what is legally mandated, \textit{and in fact must go beyond what is ``comfortable"} in order to fully realize a field that is equitable, inclusive, and diverse in identities, ideas, and lived experiences. 


\clearpage
\begin{center}
    {\large Acknowledgements}
\end{center}
This work was strongly supported in text and themes by Letters of Interest submitted to the Snowmass2021 Planning Process. These letters are listed here (by author-order): 

\begin{itemize}
    \item Assamagan, K.A., et al. (2020a) ‘Climate of the field.’ \cite{LOI_ClimateGeneral}
    \item Assamagan, K.A., et al. (2020b) ‘Educational resources for the community.’ \cite{LOI_EducationalResources}
    \item Assamagan, K.A., et al. (2020c) ‘Recruitment, evaluation, and recognition.’ \cite{LOI_RecruitmentAndRecognition}
    \item Back, A., et al. (2020) ‘Community support for enforcing collaboration codes of conduct.’ \cite{LOI_CodesOfConduct}
    \item Bard, D. and Bellis, M. (2020) ‘The effect of HEP software culture and practices on diversity, inclusion, and retention.’ \cite{LOI_HEPSoftware}
    \item Buuck, M., et al. (2020) ‘Creating inclusive collaborations.’ \cite{LOI_LZ}
    \item Nord, B., et al. (2020) ‘Culture change is necessary, and it requires strategic planning.’ \cite{LOI_StrategicPlanning}
    \item Palladino, K. (2020) ‘A need for alternative collaborative means to address misconduct.’\cite{LOI_AlternativeMeans}
    \item Pronskikh, V. and Meehan, S. (2020) ‘Engineering roles and identities in the scientific community: towards participatory justice.’ \cite{LOI_EngineeringRoles}
\end{itemize}

To that end, we would like to thank the following people for their contributions: K\'et\'evi A. Assamagan, Ashley Back, Robert Bernstein, Carla Bonifazi, Johan S. Bonilla, Daniel Bowring, Micha Buuck, Mu-Chun Chen, Michelle J. Dolinski, Alex Drlica-Wagner, Kirsty Duffy, Bo Jayatilaka, Alvine Kamaha, L. J.  Kaufman, Michael Kirby, Naoko Kurahashi Neilson, Cecilia Levy, Kendall Mahn, Rachel Mannino, Sam Meehan, Mark D. Messier, Brian Nord, Kimberly Palladino, Thomas Rainbolt, Bryan Ramson, Kate Scholberg, Sara Simon, Erica Snider, Jason St. John, Kelly Stifter, Jon Urheim, Justin Vasel, Tammy Walton, Lindley Winslow, and Jeremy Wolcott.

\clearpage
\renewcommand*{\bibfont}{\small}
\setlength{\bibsep}{4pt}
\bibliographystyle{unsrt}
\bibliography{MAIN}



\end{document}

%% file: SummaryOfRecs.tex
We summarize here the recommendations compiled as a result of this work, based on the recommendations provided by Nord \textit{et al.} in the ``Culture change is necessary, and it requires strategic planning" letter of intent submitted as part of the Snowmass planning process\cite{LOI_StrategicPlanning}.  

\begin{center}
    \subsection*{Recommendations for Funding Agencies}
\end{center}

\renewcommand{\arraystretch}{1.5}
\begin{tabular}{p{.05\textwidth} p{.05\textwidth} p{.75\textwidth}}

\textbf{1.0} & & \textbf{HEPA communities must employ the use of robust strategic planning procedures, including a full re-envisioning of science workplace norms and culture.} \\
    & F1.1 & Prioritization of climate-related issues at the funding level. This might include the inclusion of climate-related topics into safety parts of collaboration ``Operational Readiness Reviews," ``Conceptual Design Reviews," or similar documentation submitted to funding agencies.\\
    & F1.2 & Funding agencies should provide formal recommendations for institutions and collaborations for reporting violations of their codes of conduct to the funding agency itself. If there is no mechanism for reporting misconduct to a funding agency, that mechanism should be developed.\\ 
    & F1.3 & For collaborations and other institutions that are not beholden to a governing body in a legal sense, funding agencies should provide formal recommendations regarding enforcement of codes of conduct, including handling community threats, removal of collaboration affiliates, and leadership rights and responsibilities, and protections against legal liability for leadership that is responsible for that enforcement.\\
    & F1.4 & Funding and structural aid should be made available to develop ``Collaboration services'' offices at host laboratories that can provide HEPA collaborations and other physics communities with the following: a) advice on topics including those listed in F1.3, b) training in project management, ombudsperson roles, and victim-centered approaches to investigations, and c) logistical tools including facilitation of investigation and mediation.
\end{tabular}

\vspace{0.3in}

\begin{tabular}{p{.05\textwidth} p{.05\textwidth} p{.75\textwidth}}
\textbf{2.0} & & \textbf{HEPA communities must implement new modes of community organizing and decision-making that promote agency and leadership from all stakeholders within the scientific community.}\\ 
    & F2.1 & Funding agencies should facilitate Climate Community Studies, instead of leaving such studies up to individual communities to complete. In line with F3.2, these studies should be informed by expertise in social and organizational dynamics.\\
\end{tabular}

\vspace{0.3in}

\begin{tabular}{p{.05\textwidth} p{.05\textwidth} p{.75\textwidth}}
\textbf{3.0} & & \textbf{HEPA communities must engage in partnership with scholars, professionals, and other experts in several disciplines, including but not limited to anti-racism, critical race theory, and social science.}\\
    & F3.1 & Funding should be made available to both engage with and compensate such experts. This can take the form of independent grants, but more effective would be the inclusion of climate-related topics into safety components of collaboration ``Operational Readiness Reviews,'' ``Conceptual Design Reviews,'' or similar documentation submitted to funding agencies. \\
    & F3.2 & In line with F2.1, community studies should receive advice from experts in sociology and organizational psychology, such that the tools used to evaluate the climate are adequate, effective, and informative. These studies and accompanying expertise should be funded at the federal and institutional levels. \\
\end{tabular}


\clearpage
\begin{center}
    \subsection*{Recommendations for the Community}
\end{center}

\renewcommand{\arraystretch}{1.5}
\begin{tabular}{p{.05\textwidth} p{.05\textwidth} p{.75\textwidth}}
\textbf{1.0} & & \textbf{HEPA communities must employ the use of robust strategic planning procedures, including a full re-envisioning of science workplace norms and culture.} \\
    & C1.1 & Institutions and HEPA communities must develop reporting mechanisms and sanctions for egregious behavior and should transparently describe those mechanisms in full for the benefit of all affiliates. Communities must be prepared to exercise those mechanisms. \\
    & C1.2 & The community should prioritize the implementation of best practices networks across institutions and communities of physics practice. This may be facilitated through Collaboration Services Offices (F1.4), but may also include the facilitation of networks between DEI groups at similar collaborations.\\  
    & C1.3 & Future HEPA community codes of conduct should align with, and current codes of conduct should be reviewed upon new recommendations from funding agencies regarding enforcement and disciplinary measures.
\end{tabular}

\vspace{0.3in}
\begin{tabular}{p{.05\textwidth} p{.05\textwidth} p{.75\textwidth}}
\textbf{2.0} & & \textbf{HEPA communities must implement new modes of community organizing and decision-making that promote agency and leadership from all stakeholders within the scientific community.}\\ 
    & C2.1 & Reviews of community climate should include an evaluation of how leadership is selected within HEPA collaborations (e.g. assignment of high-impact analyses \& theses topics, convenership of working groups, public-facing roles representing the collaboration such as spokespersons or analysis announcement seminars)\\
    & C2.2 & Reviews of community climate should include the valuation of sub-community contributions. This includes the participation of ``non-scientists" in community engagement and authorship, community perceptions of operations and service work, the development of onboarding and early-career networks, and implementation of policies toward equity in information sharing and software. \\
\end{tabular}

\vspace{0.3in}

\begin{tabular}{p{.05\textwidth} p{.05\textwidth} p{.75\textwidth}}
\textbf{3.0} & & \textbf{HEPA communities must engage in partnership with scholars, professionals, and other experts in several disciplines, including but not limited to anti-racism, critical race theory, and social science.}\\
    & C3.1 & Experts should be adequately integrated into HEPA communities, including collaborations, such that their expertise can be applied effectively. This may take the form of an official collaboration role like a non-voting member of a collaboration council. \\
\end{tabular}

%% file: introduction.tex
Snowmass as a whole stresses the importance of and value that high energy physics \& astrophysics (HEPA) brings to our communities, local and global. The importance of investing in \emph{people} cannot be overstated. Racism, misogyny, and other fundamentally oppressive, violent, and exclusionary ideas and practices are systemic and permeate society. Scientific research environments --- e.g., academic departments, national laboratories, and collaborations --- are no exception in this regard: actions and patterns of behavior by individuals and groups, as well as policies and procedures are among the vectors of oppression. 

 The field of physics has largely failed to provide equitable opportunity to all who have the desire to perform physics research. This is evident in a number of ways, perhaps most clearly in the extreme underrepresentation of gender and racial identity, and sexual orientation in the physics community\cite{porter-AIPStatisticalResearchCEnter-WomenPhysics-2019, ivie--AfricanAmericans-2014,LGBTQ-climate-IOP,atherton--LGBTClimate-2016}. The evidence also includes the underrepresentation of Black scientists and other scientists of color, regular acts of harassment against scientists of color (including women and people of minoritized genders), and systematic disenfranchisement of marginalized and minoritized people throughout pathways and systems of education\cite{TEAMUPreport2020}. 
 
 Since 1995, the percentage of bachelor's degrees earned by African-American students has more than doubled. The overall number of bachelor's degrees awarded in physics is also the highest it has ever been. Yet, across several fields, physics has shown the largest decrease in bachelor's degrees earned by African-American students\cite{TEAMUPreport2020}.
 
 A 2018 report from the National Academies of Sciences, Engineering, and Medicine (NASEM) on Sexual Harassment of Women stated that 58\% of women academic faculty and staff and 20-50\% of women students experienced sexual harassment \cite{nasem2018}. A 2017 survey of undergraduate women in physics conducted at APS CUWiP reported 74.8\% had experienced at least one form of sexual harassment \cite{aycock2019}.
 
 \begin{quote}
“I teach a physics class. I have 30 students, half of them are women, 5 or 6 of them [will experience sexual or gender harassment] before they graduate. These are my students.... This isn’t abstract, these are students who I have been charged with educating who are experiencing this.”
\begin{flushright}
    Mark Messier, EDI Chair for the NOvA collaboration and \\ Distinguished Professor of Physics at Indiana University
\end{flushright}
\end{quote}
 
 A 2017 study of 474 astronomers and planetary scientists showed that 35\% of women of color had experienced verbal harassment related to their race, and 44\% of women of color and 43\% of white women experienced verbal harassment related to gender. Additionally, 18\% of women of color and 12\% of white women reported skipping at least one professional event because they did not feel safe attending, a huge loss of professional opportunities due to hostile climate\footnote{Averaging across all degree levels, women make up about 20\% of physics\cite{porter-AIPStatisticalResearchCEnter-WomenPhysics-2019}. At the time of this writing, there are 4334 members of the Snowmass Slack. It follows that there are approximately 870 women involved in the Snowmass process. Even if half of these were even moderately involved, following the numbers from the 2017 study \cite{clancy-JGeophysResPlanets-Doublejeopardy-2017}, more than 60 women may skip Snowmass events as a result of not feeling safe at them. This is the equivalent of the entire Axion Dark Matter eXperiment (ADMX) collaboration skipping a conference because they fear for their safety. We find that unacceptable.

}\cite{clancy-JGeophysResPlanets-Doublejeopardy-2017}
. This study explicitly focused on events that occurred between 2011-2015, indicating that these experiences are not indicative of a time long past but recent occurrences of racial and gender (including sexual) harassment.

A recent survey of LGBT+ physicists by the American Physical Society's Ad Hoc Committee on LGBT+ Issues found that over 50\% of gender-nonconforming physicists and over 50\% of transgender physicists have observed exclusionary behavior, and nearly 50\% of transgender physicists have \textit{experienced} exclusionary behavior\cite{atherton--LGBTClimate-2016}. Women respondents experienced exclusionary behavior at three times more often than men.

This problem exists in society at large, but as physicists, we have a particular ability and obligation to address inequities in our field. This is not to say that efforts to address these problems are completely lacking. Occasionally, the realities of (violence and oppression experienced every day by) minoritized people have punctured the relatively quiescent spheres of research culture and communities. These communities have traditionally responded in a variety of ways --- diversity and inclusion programming, short and intense efforts to brainstorm solutions, and long-running discussions. In fact, recent efforts to improve the gender balance of undergraduate physics education have shown some promise \cite{TEAMUPreport2020}, but these benefits do not extend to all marginalized groups. 

Unfortunately, modern concepts and practices performed under the banner of `equity,’ `diversity,’ and `inclusion’ often fall far short of directly addressing injustice, and cannot be entrusted with driving the process for change \cite{prescodweinstein-Space+Anthropology-DiversityDangerous-2018}. Additionally, traditional decision-making practices that emphasize hierarchy and one-way information flow have further embedded white supremacy within scientific communities. 

It is imperative that all of physics, and for the purposes of this paper, specifically institutions and physics collaborations, incorporate best practices to make our scientific environments spaces where all members feel safe to share their ideas, and are able to efficiently and productively contribute to the success of their experiments. While this paper spends some time focusing on collaborations as an example of an area where the community can act directly, community and collaboration action, institutional policies, and calling on funding agencies to prioritize the well-being of our community are all required and in many cases should follow similar principles. Effective solutions require action at all these levels. 

Creating plans for culture change and implementing solutions without deeply understanding the problems and without engaging experts in racism and related systems of oppression is likely to lead to more harm to researchers who are already marginalized and minoritized in science. We need to reform or remove these structures to enable lasting change. Such reform should be \textit{incentivized} through strategic initiatives and through systems that already provide existing grants. 

\begin{quote}\begin{center}
\textbf{We must remember that problems of inequity in our field will not solve themselves, and that we will have to confront them head-on, every day, likely well into the future.} 
\end{center}\end{quote}


Furthermore, there is rising awareness of the need for action to improve diversity, equity and inclusion practices in the physics community writ large both as a fundamental right and to support an effective workforce. Addressing barriers to this workforce is integral to a healthy science and technology program, and it is commonplace for institutions and physics communities to lean on this rationale that diversity is ``instrumentally useful" \cite{starckHowUniversityDiversity2021}. However, this perspective decenters the benefits to minoritized populations and centers the objectives of the institution. 
As referenced by our introductory quote: 

\begin{quote}\begin{center}
\textbf{Addressing inequities and injustices makes for better physics. \\More than that, it is simply the right thing to do.}
\end{center}\end{quote}

\subsection{Active Steps Towards Improving the Climate of the Field}
    
Improving the climate requires a multi-faceted approach that addresses both formal aspects surrounding climate as well as the broader cultural sentiments of the community. Lasting change will require fundamental organizational and cultural change that is under-girded by research and application of expertise, along with sustained, multi-generational effort within the scientific community. Moreover, these changes should be constructed and organized directly under the goal of justice. Recently, several groups have undertaken efforts based on research, scholarship, and other modes of long-developed experience and expertise in anti-racism, justice, and theories of change. For example, the recent Academic Strike For Black Lives and Shutdown of STEM\cite{ParticlesForJusticeWebsite} and Academia used community organizing practices to draw attention to the problem and to call for real\cite{ShutDownSTEMWebsite} action. More institution-oriented efforts, such as APS-IDEA \cite{APSInclusionWebsite}, are engaging theories of change and other (e.g. \cite{taplin--TheoryChange-2012}) organizational tools to envision and implement second-order culture changes in scientific research environments. Many groups of early-career researchers across the United States also responded to the oppressive systems in their communities with calls for action within their academic environments (e.g. \cite{IdeaJusticeWebsite}). Change-Now\cite{ChangeNowWebsite}, another example, established a vision and a mission for an anti-racist workplace at Fermi National Accelerator Laboratory (Fermilab). Within that context, Change-Now called for specific actions --- including organizational restructuring and policy implementations. These calls for action are supported by research and writing from scholars who bridge the physical sciences, science and technology studies, critical race theory, women’s studies, and other areas (e.g. \cite{prescodweinstein-SignsJournalofWomeninCultureandSociety-MakingBlack-2020, Hodari-2011,clancy-JGeophysResPlanets-Doublejeopardy-2017,isler-TheNewYorkTimes-OpinionBenefits-2015}). To this end, this work supports the recommendations provided by Nord \textit{et al.} in the ``Culture change is necessary, and it requires
strategic planning" letter of intent submitted as part of the Snowmass planning process\cite{LOI_StrategicPlanning}; these recommendations are as follows: 

\vspace{.3in}
\recommendation{1.0}{HEPA communities must employ the use of robust strategic planning procedures, including a full re-envisioning of science workplace norms and culture.}

\recommendation{2.0}{HEPA communities must implement new modes of community organizing and decision-making that promote agency and leadership from all stakeholders within the scientific community.}

\recommendation{3.0}{HEPA communities must engage in partnership with scholars, professionals, and other experts in several disciplines, including but not limited to anti-racism, critical race theory, and social science.}

While some systemic elements may require more time than others to fully change, we can begin to address every issue now --- establish goals and priorities, and initiate the sustained action and habits that will be required to make our research environments just. This process should also prioritize the inclusion, leadership, and decision-making of stakeholders who are most directly affected by the oppressive systems. Finally, energy, creativity, and well-understood protocols are used for planning and executing long-term programs and experiments to make new discoveries. We call on the community to engage in culture change with similar imagination and discipline.

\subsection{Overview of This Work}

We begin in Section \ref{sec:OtherReports} with reports from other STEM communities, including reports from the National Academies of Sciences, Engineering, and Medicine, the Fusion Energy Sciences Advisory Committee (FESAC), and the American Institute of Physics, among others; these recommendations should be considered alongside those presented in this work. Section \ref{sec:Institutions} describes current holes in institution policy and questions that need to be answered by funding agencies to facilitate justice and equity at host laboratories. Section \ref{sec:Collaborations} describes the work currently ongoing in a selection of HEPA collaborations, as well as the need for guidance in developing collaboration-level policies. 

Turning away from formal policies, section \ref{sec:EngineersAndScientists} discusses the need for cultural connections between hierarchies of power in HEPA, with specific focus on the disconnects between engineers and scientists. Section \ref{sec:InfoSharingHEPSoftware} discusses advancements in equitable information sharing and development/distribution of HEP software.  Section \ref{sec:NonCodeTopics} touches on a variety of topics that should be addressed over the next ten years --- namely the compensation of expertise, diversity in leadership, community networks, and organizational culture for DEI. Finally we close with a discussion on the development of a network at host laboratories to facilitate changes in institution and collaboration climates, in section \ref{sec:CollaborationServices}.

%% file: OtherClimateReports.tex

%


 The HEPA community is behind the curve and needs to act when it comes to community and collaboration action, institutional policies, and calling on funding agencies to prioritize the well-being of our community. Several recent reports have highlighted this need and begun to outline plans for action. The National Science Board Report\cite{NationalScienceBoard2020} outlined barriers to participation and advocated for solutions. Recent APS, AIP, and IOP reports have detailed impacts on communities including under-represented minorities, women, and LGBTQ+ (SGM) groups \cite{TEAMUPreport2020,LGBTQ-climate-IOP,atherton--LGBTClimate-2016,nasem2018}. The Fusion Energy Sciences Long Range Plans\cite{PoweringFutureFusion2021} and National Academies Plasma Decadal\cite{CommunityPlanFusion2020} and Astro Decadal\cite{nationalacademiesofsciencesPathwaysDiscoveryAstronomy2021} reports have included significant sections regarding their plans for action. This should be part of ongoing high energy physics activities, and our community should incorporate and build on this work. While this paper focuses on institutions and HEPA collaborations, these issues apply broadly. While we could not cover each of the well articulated recommendations in these reports, \textbf{we encourage the community to consider the following reports and recommendations for making marked improvements to the climate of HEPA.}

\subsection{Fusion Energy Sciences Advisory Committee (FESAC)}
The APS Division of Plasma Physics completed their Community Planning Process (or CPP, analagous to the Snowmass Process)\cite{CommunityPlanFusion2020} which developed concepts for improved inclusion at the community level (as this paper is part of in Snowmass). The report leads with a statement at the Executive Summary level which recognizes that having a healthy climate of diversity, equity, and inclusion is essential to the field, acknowledging the current unhealthy climate, and committing to pursuing improvement.  This should be considered for Snowmass.  Its report discussed the following recommendations, which we support and discuss in following sections of this paper. First, the onboarding of subject matter experts in issues relating to inclusion, diversity, and climate (section \ref{sec:ExpertsAndExpertise}). Such experts may spearhead the design and development of tools to assess the climate of the community on a regular basis, and can provide expert recommendations which should be implemented. Second, development and publication of new policies and codes of conduct which apply to and are reiterated at all technical meetings.  

It makes further recommendations that are not in the scope of this paper but should be developed in our community, including as part of related papers in Snowmass. These include: 
\begin{itemize}
    \item[a)] Incorporate consideration and promotion of Diversity, Equity and Inclusivity efforts as an integral aspect of the review process for institutions seeking federal funding, such as by adding a review criteria and/or DEI statement in analogy to a data management statement in the ``Program Policy Factors" (DOE) or ``Broader Impacts" (NSF);
    \item[b)] Create an accessible environment for all members of our community including those with disabilities;
    \item[c)] Increase funding opportunities for underrepresented groups including expansion of recruiting and consideration of funding mechanisms;
    \item[d)] Create parental leave and support policies, including pay and adjustment of project milestones as well as availability of lactation facilities.
\end{itemize}
 Related to (d) and particularly relevant to the current environment, flexible work and telecommuting options should be increased. Recommendations on workforce development and outreach are also important to create pathways into our science areas that help to mitigate barriers and create opportunities for under-represented groups.  The APS Bridge program is one such example.  The CPP report additionally identifies productive paths including increasing the flexibility of educational options that lead into the field to mitigate losses for the less privileged, and creation of domain-specific education and summer internship programs that can draw less represented populations into the field.

The Fusion Energy Sciences Advisory Committee Long-Range Plan report in 2021\cite{PoweringFutureFusion2021} followed the APS DPP and is analogous to P5 for HEPA. It included the following as high-level recommendations:
    \begin{enumerate}
    \item \begin{quote}
        ``Recommendation: DOE and FES should develop and implement plans to increase diversity, equity, and inclusion (DEI) in our community. Done in consultation with DEI experts and in collaboration with other institutions, this should involve the study of workplace climate, policies, and practices, via assessment metrics and standard practices."
            \end{quote}
    
        \item \begin{quote}``Recommendation: Restore DOE’s ability to execute discipline-specific workforce
        development programs that can help recruit diverse new talent to FES-supported fields of research"
            \end{quote}
    \end{enumerate}
The report includes both top-level recommendations and a full appendix devoted to detailing issues and actions, which is a structure that should be mirrored in our community. The report acknowledges current issues of representation and climate, and the barriers these pose. The committee further specifically addressed the lack of tools to both recruit and, more importantly, retain diverse talent due to systemic issues with the community culture and funding mechanisms. It advocates for the creation of new programs including workforce development, internships, training, fellowships and faculty development to increase opportunities for women, underrepresented minorities, and other underrepresented groups. The FESAC committee also called for bias and cultural competency training for program managers and primary investigators of funded projects, which should (again) be driven by subject matter experts; one example of such expertise could come from recommendations from \cite{nasem2018}.  It advocated for policies to reduce the impact of implicit bias including double blind review, and for work-life balance policies and incorporation of DEI and workforce considerations into the proposal review process similar to those detailed above. As an agency specific report, it identified actions that can be taken by the agency and others that can be taken in collaboration with other agencies and stakeholders.  The later included workplace environment and code of conduct policies, training, creation of a welcoming and accessible environment, and flexible parental leave and educational options.


\subsection{National Academies of Sciences, Engineering, and Medicine}
The National Academies of Sciences, Engineering, and Medicine have released multiple consensus reports that have taken a systemic look at addressing harassment and discrimination.  Targeted reports include: \textit{Sexual Harassment of Women: Climate, Culture, and Consequences in Academic Sciences, Engineering, and Medicine}; \textit{Graduate STEM Education for the 21st Century}; \textit{The Science of Effective Mentorship in STEMM};  \textit{Minority Serving Institutions: America’s Underutilized Resource for Strengthening the STEM Workforce}; \textit{National Academies workshop report on The Impacts of Racism and Bias on Black People Pursuing Careers in Science, Engineering, and Medicine: Proceedings of a Workshop, 2020}; and \textit{National Academies Promising Practices for Addressing the Underrepresentation of Women in Science, Engineering, and Medicine: Opening Doors, 2020}.  Such studies need further development to understand and develop practices for the mitigation of discrimination and related problems. Crucially, discrimination includes both differential treatment (including harassment) on the basis of identities and ostensibly neutral practices that produce differential impacts owing to identity.  These reports indicate that while the legal framework is essential, proactive efforts are needed to create diverse, inclusive and equitable environments. As one example the National Academies report on Sexual Harassment of Women notes that: “Four aspects of the science, engineering, and medicine academic workplace tend to silence targets of harassment as well as limit career opportunities for both targets and bystanders: (1) the dependence on advisors and mentors for career advancement; (2) the system of meritocracy that does not account for the declines in productivity and morale as a result of sexual harassment; (3) the ‘macho’ culture in some fields; and (4) the informal communications network, through which rumors and accusations are spread within and across specialized programs and fields.”\cite{nasem2018} 

    \begin{quote} 
    ``Five features in particular, especially in combination, are found to be most predictive of toxic workplace environments: (1) a perceived tolerance for harassment or discrimination, which is the most potent predictor of these occurring in an organization; (2) male-dominated work settings in which men are in positions of authority—as deans, department chairs, principal investigators, and dissertation advisors—and women are in subordinate positions as early-career faculty, graduate students, and postdocs; (3) environments in which the power structure of an organization is hierarchical with strong dependencies on those at higher levels; (4) a focus on “symbolic compliance” with federal laws that should have teeth if properly implemented --- especially Title IX and Title VII --- resulting in policies and procedures that protect the liability of the institution but are not effective in preventing harassment and discrimination; and (5) leadership that lacks the intentionality and focus to take the bold and aggressive measures needed to reduce and eliminate harassment and discrimination."
    
    \hfill \textit{on the NASEM reports} \cite{nationalacademiesofsciencesPathwaysDiscoveryAstronomy2021}
    \end{quote}


\subsection{National Academies Decadal Survey on Astronomy and Astrophysics 2020}
More recently, these and other reports have been taken into account in National Academies disciplinary scientific reports, beginning the process of analyzing and proposing actions to address issues in their research communities. These include the National Academies Decadal Survey on Astronomy and Astrophysics 2020, titled ``Pathways to Discovery in Astronomy and Astrophysics for the 2020s"\cite{nationalacademiesofsciencesPathwaysDiscoveryAstronomy2021}. 

The report makes perfectly clear that a diverse and inclusive workforce needs to be a long-term priority for the field, and acceptance of discrimination, racism, bias, and harassment, will hamper the progress of astronomy and astrophysics. The authors state that discrimination and harassment lead to negative impacts on personal and professional well-being, equitable participation, and economic prosperity. They specifically call out the injustice in watching perpetrators of harassment and discrimination be at best, tolerated, and at worst, professionally rewarded. They call for a zero tolerance policy for those who abuse either their position, their colleagues, or both. 

\begin{quote}
    What needs to be done to fully address this issue once and for all is not a mystery; there are well-established best practices, documented solutions, and veritable how-to guides that can be implemented at the individual, organizational, and profession-wide levels... that the astronomy and astrophysics community could endorce, adopt, and most importantly, work deliberately to implement. These include an especially important role for the federal funding agencies that, backed by existing federal laws, can use the power of the purse as a forcing function to help drive needed change.
    
    \hfill Pathways to Discovery in Astronomy and Astrophysics for the 2020s \cite{nationalacademiesofsciencesPathwaysDiscoveryAstronomy2021}
\end{quote}

\begin{quote}
    Recommendation: NASA, NSF, DOE, and professional societies should ensure that their scientific integrity policies address harassment and discrimination by individuals as forms of research/scientific misconduct.
    
    \hfill Pathways to Discovery in Astronomy and Astrophysics for the 2020s \cite{nationalacademiesofsciencesPathwaysDiscoveryAstronomy2021}
\end{quote}

\paragraph{Demographics and Survey Data} The AstroDecadal report also discusses the collection of demographic data as a necessity for assessment of strategies toward a more diverse and inclusive workforce. They point out that funding agencies do not collect the same quantity or categories of demographic data, and much of what is collected is not accessible or designed for data analysis. The report requests that funding agencies implement a working group for ``establishing a consistent format and policy for regularly collecting, evaluating, and publically reporting demographic data and indicators pertaining at a minimum to outcomes of proposal competitions" \cite{nationalacademiesofsciencesPathwaysDiscoveryAstronomy2021}. This enables further understanding of the solutions for addressing diversity and inclusion, and leads directly into the last recommendation: 

\begin{quote}
    Recommendation: NASA, DOE, and NSF should consider including diversity --- of project teams and participants --- in the evaluation of funding awards to individual investigators, project and mission teams, and third-party organizations that manage facilities.
    
    \hfill Pathways to Discovery in Astronomy and Astrophysics for the 2020s \cite{nationalacademiesofsciencesPathwaysDiscoveryAstronomy2021}
\end{quote}




\subsection{The AIP National Task Force to Elevate African American Representation in Undergraduate Physics \& Astronomy (TEAM-UP)}
In 2019, the AIP National Task Force to Elevate African American Representation in Undergraduate Physics \& Astronomy (TEAM-UP) released their report titled ``THE TIME IS NOW: Systemic Changes to Increase African Americans with Bachelor’s Degrees in Physics and Astronomy"\cite{TEAMUPreport2020}. Their two-year study, which included surveys, interviews, and site visits, ``identified five factors responsible for the success or failure of African American\footnote{The authors of the TEAM-UP report identify that the terms \textit{African American} and \textit{Black} may hold different meanings for different community members. The authors of this work elect to use the term \textit{African American} in this section in line with the chosen language of the TEAM-UP report.} students in physics and astronomy: Belonging, Physics Identity, Academic Support, Personal Support, and Leadership and Structures" and posed an additional five main recommendations. 

The report details how ``sense of belonging", the feeling of being part of a broader community, is an integral part of the success of African American students. Connections are easily drawn between inclusive behavior and student persistence in physics. Respondents specifically pointed out both positive and negative experiences with faculty and/or peers (including assumptions of incompetence, stereotype threat, isolation, and disparaging remarks) and drew direct lines to their sense of belonging within the physics department. The authors specifically identify that members of the community ``with social power" must actively foster engagement and sense of belonging within their communities, and call out behavior that excludes African American students. 

The development of a ``physics identity", or the perception of a community member with respect to physics, is also a strong predictor of persistence in physics. Part of this identity development involves traversing a community that is white, masculine, cisgender, and heteronormative, and overcoming the ingrained stereotypes of who can be a physicist. This identity is challenged when, as students reported, senior community members like faculty suggest that students who struggle should leave the field. Instead, the report suggests that communities examine their own assumptions about who can ``do physics" and to explore the systematic outcomes of their young members disaggregated by race, ethnicity, and gender\footnote{The authors of this work also suggest doing so by sexual \& gender identities and disability status, insofar as junior members are willing to disclose those identities.}. The report also makes mention of a diversity in leadership with respect to race and ethnicity and other identity groups in such a way that students can identify with ``people who do physics" across multiple identity groups. Communities should engage in and reward formal and informal mentoring of minoritized members, including by mentors who may not identify with a minoritized identity. The report calls on these communities to learn from experts in the field of physics education research, especially those who have successfully implemented interventions on this subject elsewhere. Finally, the authors of the TEAM-UP report make it clear that identity and lived experience are part of what makes each of our community members good at what they do; allowing members to engage with the community as their full, authentic selves is what will make our field all the stronger. 

While we heartily support the implementation of each of the TEAM-UP task force's recommendations, we list here a selection of recommendations (verbatim from \cite{TEAMUPreport2020}) that we highlight further in this work. We also suggest that these recommendations are applied beyond departments to the wider community and HEPA collaborations, which are similar communities of practice. 
\begin{center}
\noindent\fbox{\parbox{.9\textwidth}{ 
    \vspace{0.1in}
    \textbf{A Selection of Recommendations from ``THE TIME IS NOW: Systemic Changes to Increase African Americans with Bachelor’s Degrees in Physics and Astronomy"\cite{TEAMUPreport2020}:}\\
    
    Recommendation 1b. In classrooms, student clubs, and common spaces, departments should establish clear rules of engagement that ensure that everyone is welcomed and valued and convey that inappropriate behavior will not be tolerated. Departments should also provide spaces and opportunities for education and ongoing discussion among faculty and students on ways to actively foster a sense of belonging and reduce barriers to inclusion.\\ 
    
    Recommendation 1d. Departments should establish and consistently communicate norms and values of respect and inclusion. They should periodically assess departmental climate with help from outside experts and should respond, as needed, with educational workshops led by experts from student affairs or other resources. \\
    
    Recommendation 2b. Departments should examine whether their current activities foster physics identity, assess their efficacy across race/ethnicity/gender and other social identities, and modify such activities as necessary.\\
    
    Recommendation 2c. Departments should diversify their faculty with respect to race/ethnicity/gender and other social identities in such a way that support of underrepresented students is provided by multiple faculty of varying identities.\\
    
    Recommendation 3b. Departments should adopt policies and practices that encourage multiple faculty, including those who are not members of marginalized groups, to engage in formal and informal mentoring of students, and they should recognize and reward these efforts.\\
    
    Recommendation 4d. Faculty should strive to understand that students do not leave behind their identity and experiences when entering the classroom and should recognize the unique promise of each student from a perspective of strengths rather than weaknesses.} } \\ \vspace{0.2in}
    
\end{center}

We leave the most important notes on this report for last: the call to action. The TEAM-UP report recommends that physics communities like those we have developed in HEPA, including collaborations, should ``each develop a theory of change utilizing sensemaking and shared leadership." We should learn from reports, like those listed in this work, and educate ourselves on the barriers to inclusion for underrepresented groups in HEPA spaces. The report calls for our organizations to engage in learning communities and skill-building workshops, to establish incentives for DEI efforts, to collect demographic and identity-based data for metrics on decision-making, to sanction misconduct and behavior that does not align with shared values, and show that ``equity and inclusion are valued." We leave you with the following: 
\begin{quote}
    This report calls for significant changes in the culture of physics and astronomy. These changes will be neither easy nor quick. The research literature suggests that second-order change in higher education takes 7 to 10 years or longer\cite{kezarHowCollegesChange2018}, and often efforts toward change do not succeed. We cannot in good conscience wait for these improvements. Students need to experience environments in which they can learn, grow, and prosper, now.
    
    \hfill ``THE TIME IS NOW: Systemic Changes to Increase African Americans with Bachelor’s Degrees in Physics and Astronomy"\cite{TEAMUPreport2020}
\end{quote}


\subsection{LGBTQ Climate Reports (APS \& IOP)} 
In 2016 an Ad-Hoc Committee on LGBT Issues released its report on ``LGBT Climate in Physics: Building an Inclusive Community" \cite{atherton--LGBTClimate-2016} which reviewed the status of and barriers faced by LGBT physicists within the physics community. They wrote that the overall climate of the field as experienced by LGBT physicists varied widely but identified several findings that we would like to highlight (verbatim from \cite{atherton--LGBTClimate-2016}):  
\begin{itemize}
    \item ``In many physics environments, social norms established expectations of closeted behavior."
    \item ``A significant fraction of LGBT physicists have experienced or observed exclusionary behavior."
    \item ``LGBT physicists with additional marginalized identities faced greater levels of discrimination" 
    \item ``Transgender and gender-nonconforming physicists encountered the most hostile environments"
\end{itemize}

Over 40\% of respondents to a survey posed by the committee agreed that ``employees are expected to not act too gay." Over 50\% of gender-nonconforming physicists and over 50\% of transgender physicists have observed exclusionary behavior, and nearly 50\% of transgender physicists have \textit{experienced} exclusionary behavior. Those who have additional marginalized identities fare worse: LGBT female physicists reported physical and verbal sexual harassment, exclusionary behavior, and discouragement to report. LGBT physicists of color reported homophobic and racist comments, stereotype threat, and lack of community. One respondent was deterred from pursuing a PhD in physics because of the overlap of her race and queer identity, specifically due to the risk of perceived inferiority. Transgender respondents reported that colleagues in their department intentionally misgendered them and mocked or openly laughed at them. They report accusations of predation and protests against their very presence on campus. Attending class or showing up to work was a hazard to their safety, and required them to constantly be on edge for confrontation or legal action. 

We support the recommendations of the ``LGBT Climate in Physics: Building an Inclusive Community" report \cite{atherton--LGBTClimate-2016}, with the understanding that physics must be a safe and open community --- free of explicit or implicit violence against those who identify as LGBT. All of the recommended actions should be included, but we would like to highlight the following text which aligns most closely with the recommendations in this work: 

\begin{center}
\noindent\fbox{\parbox{.9\textwidth}{ 
    \vspace{0.1in}
    \textbf{A Selection of Recommendations from ``LGBT Climate in Physics: Building an Inclusive Community" \cite{atherton--LGBTClimate-2016}:}\\
    
    \textbf{APS should implement the Code of Conduct with thorough and careful regard to informing members and responding to reports of infractions}.... The APS staff attending meetings should also be trained to not challenge the validity or severity of any claim. APS staff should then see to it that the participant is connected as soon as possible with designated staff members who are more specifically trained to address the alleged infractions.... APS should establish protocols regarding the handling of infractions of the Code of Conduct. A process should be established to formally document all reports to the APS. Further, the Code of Conduct should be enforceable. We encourage APS to clarify the procedures it will follow to establish the veracity of alleged infractions, as well as actions it is prepared to take if an infraction is verified to have occurred. \\
    
    \textbf{APS should lobby federal funding agencies to include LGBT demographics in STEM education and workforce surveys and to acknowledge a pressing need to address climate issues for LGBT people in STEM fields}.... Federal funding agencies value efforts to promote women and racial or ethnic minorities in STEM fields as addressing the “broader impacts” review criterion used to evaluate proposals. Similar efforts relevant to LGBT individuals in STEM fields are typically not included. Given the evidence regarding climate issues and exclusionary behavior faced by LGBT individuals in physics, as well as science and engineering on a broader scale. APS should lobby funding agencies to promote LGBT inclusion and mitigate discrimination. APS should value such efforts as addressing the broader impacts review criterion.\\
    
    \textbf{APS should develop a training program on inclusive workplace and mentorship practices for physicists in academia, national labs, and industry that incorporates the needs of LGBT physicists and aims at the recruiting of active allies}.... APS can take a number of steps to foster active allies and provide information on LGBT inclusion. APS should provide resources and training to aid in the professional development of active physicist allies whose efforts will ultimately result in the physics community adopting more inclusive practices.

    } } 
    
    \vspace{0.2in}
    
\end{center}

A 2019 report by the Institute of Physics, Royal Astronomical Society and Royal Society of Chemistry titled ``Exploring the workplace for LGBT+ physical scientists" \cite{LGBTQ-climate-IOP} had similar recommendations. In a 2018 survey, 28\% of LGBT+ respondents reported that they had considered leaving their workplace due to the climate, and nearly 50\% of transgender respondents reported the same, with 20\% reporting that they thought of this often. The international nature of the survey also allowed respondents to comment on working with people from other cultures. One respondent reported that he felt the need to ``return to the closet" in order to protect a collaboration with a culture that considered his sexuality shocking; he asked that communities ``do not leave it to individuals to navigate their identity through on their own with regards to international collaborators" \cite{LGBTQ-climate-IOP}. 

The highlighted recommendations from this report are as follows, and are broken down in the text in terms of actions to be taken by individual scientists, by employers, and by the community at large.  

\begin{center}
\noindent\fbox{\parbox{.9\textwidth}{ 
    \vspace{0.1in}
    \textbf{A Selection of Recommendations from ``Exploring the workplace for LGBT+ physical scientists" \cite{LGBTQ-climate-IOP}:}\\
    
    \textbf{4.1 Building a visibly welcoming community}.... For many the culmination of good visibility was to have senior leaders and managers proactively acting as champions, whether LGBT+ or allies. A lack of senior roles held by openly out LGBT+ staff and an overall lack of obvious support from those in senior positions is noticeable. If decision makers are not effectively considering LGBT+ needs, then it seems unlikely that their policies and procedures will do so either.\\
    
    \textbf{4.2 Reviewing and improving policies}....Creating an environment where everyone feels welcome means addressing discrimination and harassment of every kind, whether this is the use of homophobic language, even when supposedly joking, or the exclusion of subgroups from LGBT+ networks. Empowering everyone to call out discriminatory actions or offensive behaviour when they see them is essential, and needs clear guidance available through management or policies. \\
    
    \textbf{4.3 Introducing and improving training}.... Training that supports LGBT+ staff could be used more effectively in nearly all work environments. Respondents would value more practical training, particularly targeting negative or harmful behaviours, such as use of homophobic language or misuse of pronouns. If carefully done, this sort of training could be invaluable to those respondents who as allies wished to call out and address this kind of behaviour.

    } } 
    
    \vspace{0.2in}
    
\end{center}




%% file: InstitutionPolicies.tex
Creating, distributing, and enforcing codes of conduct has not been pursued wholeheartedly within the particle physics community, but has in other topical communities such as astrophysics\cite{AASCodeEthics}. They have been enacted to varying degrees by individual institutions but are seen by many individuals as being documents of token-respect implemented in an umbrella way to shield the institution. Moreover, because our research is largely done in a cross-institutional manner, the enforcement of codes of conduct is unclear, particularly given the ambiguous legal relationship of many collaborations to collaboration affiliates.

If institutions do not have the means or neglect to enforce sanctions for severe violations of the code of conduct, this can lead to additional harm via institutional betrayal; this refers to wrongdoings perpetrated by an institution upon individuals dependent on that institution, including failure to prevent or respond supportively to wrongdoings (e.g. sexual assault) committed within the context of the institution \cite{platt2009}. Institutions who cannot support those reporting misconduct have the power to cause additional harm to survivors; for example, women who report institutional betrayal surrounding an experience of sexual harassment or assault have more trauma-related symptoms than those reporting in a supportive environment \cite{smith2013}. When codes of conduct are unenforceable, collaborations become an unsupportive environment.

Furthermore, the greatest predictor of the occurrence of sexual harassment is organizational climate, evaluated on three elements: (1) perceived risk to those who report sexual harassment, (2) a lack of sanctions against offenders, and (3) the perception that a report of sexual harassment will not be taken seriously \cite{nasem2018}. Without the ability to enact sanctions, we risk our scientific institutions, including HEPA collaborations, being an environment in which sexual harassment flourishes despite the existence of a code of conduct.

We have the opportunity to address recommendations from the NASEM report -- institutions moving beyond legal compliance to address culture and climate, and improving transparency and accountability -- with potential disciplinary actions in codes of conduct. We must develop community-wide policies that go beyond symbolic legal compliance, which protect institutions from legal liability but do not prevent harassment. Our institutions must develop ``clear, accessible, and consistent policies'' on harassment and standards of behavior which should include ``a range of clearly stated, appropriate, and escalating disciplinary consequences'' for those found to have violated policies and/or law \cite{nasem2018}, and these policies should be consistent across subfields and experiments.

\subsection{Handling Misconduct}

While the majority of work should be focused on preventing bullying or harassment in the first place\cite{clancy:2020}, we must be prepared to act upon the values and policies stated in our codes of conduct.

\paragraph{Is it a violation?}
Often situations arise where misconduct, such as bullying or harassment, is of a threshold nature. Interpersonal disputes may be dominated by a small number of individuals, or may not rise to the level of violating the explicit nature of the code of conduct, regardless of whether it is affecting collaborative work. To fill the void of inaction, ``whisper networks" arise --- often between junior or otherwise minoritized groups --- to attempt protection of vulnerable members of the collaboration from poor actors. When a pattern emerges, when a ``whisper network" is developed, when does that behavior rise to the ``threshold" of action? 

Whisper networks, while undeniably effective to some degree for those involved in them, are by their nature exclusionary.  A whisper network is an informal and secretive means of communicating protective information, but those who participate in them have to be careful in many instances, since including the wrong person can expose the group to retaliation. So, people with cross-sectional identities may be excluded from the protective circle of a whisper network and thus not have access to the warnings that are distributed to others, while still potentially being targeted by known threats.  Someone who is for any reason less connected to the scientific community will be less protected by nature of having fewer potential contacts with relevant information for them. This may be especially relevant for women of color, transfer students, early career individuals, or anyone with a less traditional academic background.  Though a community with a positive culture will not have the need for whisper networks, banning them outright is akin to banning stitches because open wounds are unpleasant.  The prevention of whisper networks and thereby their unequal application comes down to addressing the root causes for them. Academic institutions, including HEPA collaborations, are responsible for creating and maintaining positive work environments for their students, employees, and visitors, including enforcement of rules on collegiality and professionalism and reprimand or removal of persons that negatively impact the culture they build and invest in. 

\begin{quote}If our current system is not serving our most vulnerable members, we need to actively work on new structures to support victims and educate the community regarding behavioral norms.
\flushright{Kimberly Palladino, \textit{A Need for Alternative Collaborative\\ Means to Address Misconduct} (Snowmass LOI)}
\end{quote}

\paragraph{Who is responsible for the reporting structure?} Interest groups, professional organizations, and scientific collaborations with member institutions from all over the world have regular in-person meetings at single participating institutions. If there is an incident at an in-person meeting, the likelihood that this happens between two community members who are not members of the host institution is high. Who is then responsible for the investigation? Under existing policies, the host institution is unlikely to investigate if neither collaboration affiliate is associated with that institution. As many institutional procedures protect the liability of the institutions \cite{nasem2018}, the victim’s and the perpetrator’s institutions may not investigate.

Some institutions have stated they are not obligated to investigate if neither the victim nor the perpetrator is an employee of the lab. Some institutions do state that they will treat issues between visitors in the same way as issues between employees, but this is not consistent between all institutions, and investigations of harassment should not depend on the generosity of a particular institution. Often, experiences with human resources and legal counsel leave the impression that their objective is to protect the institution, and that resolutions to such cases prioritize the institution over victims\cite{peirce1998,greenfield2017,trump-OtagoDailyTimesOnlineNews-Failuresracism-2022, panzerSexualHarassmentProcess2019, UnderstandingIntersectionTitle2015, novacicStudentsAccusedSexual2019, andersonDeadlineTimeNew2020} --- this is happening \emph{in our institutions, in physics and astronomy}\cite{martinTitleIXInaction2020}. In many instances, this may mean that it is best, from the standpoint of institution representatives, for the institution not to handle the case.

Assuming that an institution \textit{is} willing to investigate, many are not equipped to handle egregious cases of harassment or assault, nor able to prevent retaliation after reporting. Victim-centered approaches seek to minimize retraumatization associated with an investigation by providing victim advocates and other services, engaging victims throughout the process, and giving victims the opportunity to be involved in forming disciplinary actions. Existing policies differ between institutions and often do not take any of these factors into consideration; this results in unbalanced investigations strongly dependent on where the incident took place. Institution-based disciplinary measures may be limited to restricting site or computing access, especially in the case where a perpetrator is not an employee. Because removal of computing privileges is, in effect, removing the ability of a community member to continue to work, institutions may be hesitant to take these steps for fear of legal retaliation. 

\subsection{Communication of Findings}
Some collaborations have indicated in their code of conduct that they will act upon a report of misconduct from another institution, e.g. from a professional society or a laboratory. However, public reports of misconduct that can be acted upon are rare; privacy concerns prevent communicating the findings of an investigation more detailed than whether the investigating institution is choosing to take disciplinary action. The absence of disciplinary action from a particular institution may only be a statement about the action the institution is willing to take and does not necessarily mean that the investigation yielded no evidence of misconduct. However, with no report with details of the investigation and no public disciplinary action taken by the investigating institution, professional organizations or scientific collaborations that do not have the capability of performing their own investigations have no way to address the misconduct in their own community. Therefore, organizations who may have the desire \textit{and need} to enact their own disciplinary measures to protect their community from further harm may simply be unable to do so.

%% file: CollaborationPolicies.tex


\recommendation{C1.1}{Institutions and HEPA communities must develop reporting mechanisms and sanctions for egregious behavior and should transparently describe those mechanisms in full for the benefit of all affiliates. Communities must be prepared to exercise those mechanisms.}

As experimental collaborations are not legal entities, they live in a gray area where there is no official oversight. No one institution, including the host institution of the experiment, is responsible for every member of the collaboration. We may look to experimental leadership, but it is not good policy to make the spokespeople of our experiments responsible for the behavior of collaboration members, especially when disciplinary actions taken by spokespeople in response to misconduct leaves them open to potential legal action - not something usually covered during spokesperson elections. Regardless, establishing norms and expectations for collegial behaviors as well as ways to seek accountability through codified documents is essential to protect the members of the community and provide a foundation for cultural change within it. 

Many collaborations have implemented codes of conduct, a necessary first step toward a safer working environment. These codes have been successful in raising awareness of the aforementioned issues impacting our colleagues and effective in addressing less severe violations of these codes. Though these codes often include potential disciplinary measures for severe violations, collaborations will be unable to follow through with these measures as they often require assistance from outside institutions which may be unwilling or unable to provide the support needed. If collaborations indicate that they will enforce sanctions for severe violations but fail to do so, regardless of the reason, this can lead to additional harm done to the target of the misconduct via institutional betrayal \cite{platt2009}. 

\subsection{Implementation in Practice}
\input{OmbudsInterviews}
\subsection{Common Themes and Pitfalls of Collaboration-Developed Codes} \label{ssec:Pitfalls-of-Collaboration-Codes}

    \begin{quote}
        If a community cannot hold the most powerful people in the community accountable to the code of conduct, it is best not to adopt a code of conduct at all. 
                
        \hfill \textit{How to Respond to Code of Conduct Reports}, Aurora \cite{Aurora:2019}
    \end{quote}
    When equity-minded efforts, namely code of conduct development and enforcement, are left up to collaborations, when they are left up to physicists, there are common themes that arise from community members. There are some issues, including the lack of systemic support, which can be addressed by institutional support systems as described in Section \ref{sec:CollaborationServices}. Other issues have direct and negative impacts on the climate of the community, and should be addressed as such.
    
    Some concerns raised during these meetings should be addressed in this work and, more formally, through advice from experts and funding agencies. We also include a list of common ``pushbacks" here, as well as some details on how collaborations could address them. Connecting to the running theme of this work, this section was developed by physicists and can be markedly improved by advice from experts. 

    
    \subsubsection{Pitfalls of Collaboration Developed Codes}
    \paragraph{Collaborations are not legal entities:} \label{sssec:Pitfall-NotLegalEntities}
        Collaborations do not exist in a legal sense. This may mean that the collaboration, as an organization, is protected from lawsuits, but it also means that the collaboration has difficulty acting in any official capacity. In addition, the inability to bring a lawsuit against the collaboration itself does not protect individuals acting on behalf of the collaboration, e.g. spokespeople or an executive board, from potential legal action as individuals. This means that most disciplinary actions that are detailed in codes of conduct cannot be enforced without opening individuals up to legal action and therefore, HEPA communities effectively \textit{cannot enforce the codes of conduct that they have adopted.} This is damaging in multiple ways: first, those who engage in misconduct cannot be sanctioned by the collaboration and their behavior will likely continue, and second, victims who believe that the community can protect them will be betrayed by empty promises. Funding agencies have access to experts on legal matters, including mandatory reporting, academic misconduct, and legal gray areas between agencies, institutions, and countries, and collaborations would benefit from advice on these matters. This is discussed in further detail in sections  \ref{sec:CollabHandlingMisconduct} and \ref{sec:CollaborationServices}.
    
    \paragraph{Lack of systemic support:} \label{sssec:Pitfall-SystemicSupport}
        It is increasingly common that an institution will have in place a performative code of conduct and/or a standing “EDI Committee” which largely signals to the members of that community as well as the outside world that there is a commitment to diversity and justice. Unfortunately, the precise role that such committees play is often ill-defined and accompanied by little to no practical power. For such committees to be effective in affecting meaningful change requires more careful organization and planning, structures of accountability, institutional power, and vociferous support from community/collaboration leadership. Section \ref{sec:OrganizationalCulture} discusses such institutional approaches and practices in further detail. A code of conduct should not be a document that checks a ‘requirements’ box. Instead, it should provide for and contribute to an active and sustained commitment for improving collaboration culture.
        
        Additionally, host institutions often do not provide resources for collaborations for development of their codes of conduct, which increases the difficulty and time taken to create and adopt a code of conduct. Host institutions also often do not provide tangible support for collaborations with respect to enforcement of their code of conduct, leaving investigations of misconduct and enforcement of disciplinary actions to individual people, which, as discussed in Sec. \ref{sssec:Pitfall-NotLegalEntities}, essentially makes collaboration codes of conduct unenforceable. While nothing prevents an individual that has been subject to disciplinary action by a host institution from filing a lawsuit against them, institutions have financial resources, general counsel, and a human resources department; all things that collaborations and individuals on those collaborations often do not have access to. Therefore, institutions should take on this responsibility, as they do benefit from the existence of the collaboration and have a responsibility to protect the scientists who are contributing to their scientific output and reputation.

    \paragraph{``Ombudsperson" / ``EDI chair" is a full time role:} \label{sssec:Pitfall-OmbudsOverload}
        Because so few people are able to fill these roles based on seniority, approval of collaboration affiliates, relative protection from potential retaliation by bad actors, and ability to take on this work in addition to their everyday work, ombudspeople end up filling multiple roles, without training or support. Only one collaboration that we have engaged with has provided ombudsperson training; other ombudspeople may seek out training independently but do not have access to collaboration funding for said training. Multiple collaborations reported that their ombudspeople perform various victim assistance services, for which they are not trained or supported in a tangible way. Professionally trained victim advocates are at risk of burnout and developing secondary trauma stress due to listening to accounts of traumatic experiences, and receive support via their occupations to counteract these risks. These risks may not be clear to those considering their collaboration's ombudsperson role or equivalent, and no such support to counteract these risks exists unless the ombudsperson seeks it out on their own time, with their own funds. 
        
        Additionally, those in these roles also tend to be responsible for all diversity-and-inclusion-related programming for the collaboration -- another job that people are professionally trained to do being done by physicists without training. With this approach, we are setting our colleagues up for burnout or for failure. Funding could be made available for the training of collaboration ombudspeople or EDI representatives on topics such as confidentiality, victim advocacy, trauma-informed investigations of misconduct, and emotional/psychological well-being; better yet, funding should be made available to hire experts for these roles.

    \paragraph{Efforts at addressing misconduct should be victim-centered:}
        Mediation should never be forced or coerced, and should never be orchestrated by collaboration members only. Mediation should be target-centered, and driven by a professional. Appropriate protections should be made to protect the identity and safety of the reporter. Additionally, apologies should not be solicited of the perpetrator nor should the reporter be asked to accept the apology or forgive the misconduct; rather, the purpose of mediation should be to resolve the misconduct such that the reporter is fully reintegrated into the collaboration community. 
        
        Harm to a collaboration affiliate is harm to the collaboration and to the community, and should be treated as such. This includes harm to the affiliate through retaliation or weaponization of the reporting process, as well as through institutional betrayal. Restricting a harasser's access to the reporter (for example, by changing the working group assignment of the harasser or the reporter) is not appropriate; the reporter is not made whole, and the harasser may move on to a new target. This response treats the report as a personal conflict rather than a threat to the entire community. In the case of in-person meetings, including collaboration meetings, Snowmass meetings, conferences, and the like, the reporter is provided a choice to attend and expose themselves to further harassment, or not to attend and lose opportunities that should be afforded them. The reporter in this case has been asked by the collaboration to be on constant guard against threats from the harasser, adding to the emotional and psychological burden of the originating incident(s). 
        
        Are there appropriate mechanisms in place to maintain the physical, psychological, and emotional well-being of collaboration affiliates who file reports with the collaboration? Are there persons built into the codes of conduct that act as victim advocates, and will remain at the victim’s side, with their well-being first-and-foremost? Do these persons represent collaboration diversity, including diversity of race, gender, disability, type-of-institution, and seniority? Are these persons offered training through official channels? Are they provided with reduction in their other responsibilities, or are they expected to take on this role on top of their day-to-day? The answer to these questions for many, if not all, collaborations currently is ``no."

        Codes of conduct deal with violations and problems. What are collaborations doing to be proactive? To encourage inclusive behavior? Issues that come forward might be structural or cultural. How do we create cultures that prevent code of conduct violations instead of requiring communities to react, often in less-than-useful ways?

    \paragraph{Commitment to DEI is an ongoing, all-inclusive process:} 
        Development of a code of conduct or DEI policies must be accompanied by an all-hands, continual process of evolution and commitment. Again, the code of conduct should not be a document that checks a ‘requirements’ box. Instead, it should provide for and contribute to an active and sustained commitment for improving collaboration culture. 
        
        The code of conduct and related documents should be introduced to new collaboration affiliates as part of a formal onboarding process. This responsibility should lie with collaboration organization and may include regular onboarding meetings. There are many benefits to this: 
            \begin{itemize}
                \item All members will have a recorded introduction to community ethics and the code of conduct. They may ask questions of community experts, and provide feedback.
                \item The responsibility is not with individual PIs, who may or may not prioritize the code to the same degree as the collaboration representatives. 
                \item Members who onboard at the same time have the opportunity to build community, especially if this is emphasized by DEI committee activities.
            \end{itemize}
        The code of conduct should be introduced at the start of collaboration activities, including working group meetings, collaboration meetings, and workshops. This should occur for both in-person and virtual meetings. The code of conduct should also be reviewed periodically to account for the evolution of language overtime, especially as it relates to demographic groups (e.g. \cite{castaniaEvolvingLanguageDiversity2003}). It should also reflect the values put forward by the collaboration, and all its representatives.  
        
        How does the code of conduct account for members that are not authors or are not traditionally thought of as collaboration affiliates? This might include technicians, engineers, IT specialists, or even EDI committee members (see section \ref{sec:EngineersAndScientists}). How are communication channels managed such that they are accessible by all affiliates who contribute to the collaboration? Do authorship rules, board membership, working group convenership, and/or event planning leave out contributions from those traditionally not thought of as collaboration affiliates? 
        
        We note that the implementation of the code of conduct for collaborations who use collaborative software like Slack has been unclear, especially when Slack does not allow users to block other users without administrative intervention\cite{antonelliYouCanBlock} (which, as we have described, may be difficult given the known issues with administration intervention to resolve reports). Collaborations will need to take care that the platforms they conduct business on have built-in protections that align with the code of conduct and community values.

    \paragraph{The knowledge already exists: } 
        HEPA communities are not the first to attempt these policies, and in fact we are certainly not the most equipped to do so. Funding should be made available to invite experts to speak on these topics. NSF has funded honorariums for guest speakers in the past, and should continue to do so.  Funding should be made available to propagate best practices throughout HEPA collaborations, which might include organizations of EDI Chair / Ombudsperson meetings between collaborations (e.g. the Multimessenger Diversity Network \cite{MultimessengerDiversityNetwork}).
        
        Funding should be allocated to hire equity and inclusivity experts, such as equity scholars and organizational psychologists, to advise on development of codes of conduct, DEI initiatives, and methods of tracking the success of policies outlined in their code of conduct. These experts should be included as collaboration affiliates, so that their expertise is taken seriously. This provides benefits on both sides: the collaboration is better served when they engage with someone they trust. (This suggestion is directly related to the success of this model from IceCube, section \ref{sec:Implementation_IceCube}, and SuperCDMS, Ssction \ref{sec:Implementation_SuperCDMS}.) We discuss this more in Section \ref{sec:ExpertsAndExpertise}.  
         
        Funding should be made available for collection of demographic information guided by experts, and funding agencies should be responsible for tracking and sharing it. Funding should also be made available for a centralized space with information on EDI, best practices, and funding. This should also include data on why collaboration codes of conduct and reporting mechanisms are necessary for collaboration culture, and other responses to common pushbacks (section \ref{sssec:Pushbacks}).

    
    \subsubsection{Questions on Policy Implementation}
    
    This section details a selection of thematic questions that arose during discussions with the collaboration representatives. We believe that these questions need answers from funding agencies, as advised by experts. 
    
    \vspace{0.1in}

    \recommendation{C1.3}{Future HEPA community codes of conduct should align with, and current codes of conduct should be reviewed upon new recommendations from funding agencies regarding enforcement and disciplinary measures.}
    
    \paragraph{Reporting to Funding Agencies:} \label{sssec:Question-ThirdPartyReporting}
    Because collaborations are not legal entities themselves, decisions made on behalf of the collaboration are on shaky legal footing. As discussed further in section \ref{sec:CollabHandlingMisconduct}, it also means that investigations made through collaboration efforts are probably not justification enough for sanctions on a collaboration affiliate.
    
    When institutions decline to investigate, and collaborations defer responsibility, all that is left as a remedy is the agencies who fund the work, and those processes are very opaque. Clarification is needed on these mechanisms for reporting violations to funding agencies. PIs and project managers are expected to report incidents to their respective funding agencies, but what about incidents that occur off-site, or are spread across multiple funded groups? Who speaks on behalf of the collaboration to the funding agencies in this case? Who is responsible for maintaining / tracking communication, and for enforcing any decisions made by funding agencies?  How are ongoing issues with no obvious intervention tracked and responded to? For example, are reports like ``I don’t want to get anyone in trouble but...'' still tracked and compiled for future reports? Are issues beyond the scope of the code of conduct still brought to the attention of the collaboration administration, institution offices, and/or funding agencies? Advice is also needed on the interaction between collaboration and leading laboratories or local institutions; trust in the reporting mechanisms is a key part of soliciting reports, but lab/university investigators are not members of the collaboration. How do we remove this barrier? Are there policies in place for collaborations to address reports they receive about collaboration affiliates from external organizations (e.g. APS)?

    \vspace{0.1in}
    \recommendation{F1.2}{Funding agencies should provide formal recommendations for institutions and collaborations for reporting violations of their codes of conduct to the funding agency itself. If there is no mechanism for reporting misconduct to a funding agency, that mechanism should be developed.}

    \subsubsection{Common Pushbacks to DEI Efforts} \label{sssec:Pushbacks}
    Unfortunately, HEPA communities do face pushbacks when it comes to DEI work. This section details some comments that are listed in literature or were experienced by the authors, with thoughts on how to respond. The authors credit Gosztyla \textit{et al.} (2021)\cite{Gosztyla:2021}, ``Responses to 10 common criticisms of anti-racism action in STEMM", for a substantial portion of this section's literature review.

    \paragraph{``Why can't we just do physics?": Refusing the responsibility} \label{sssec:Pushback-RefusingResponsibility}
    This is an easy argument to make if you assume that there are other people doing the work and that there are other people getting paid a fair wage to solve the problems of hierarchical structures in HEPA. The problem is: \textit{there aren't}. At least not to the extent that we need. We \textit{need} scholars and experts to be embedded in our communities. 
    
    This is also an easy argument to make if one assumes that ``since I'm not contributing to a negative climate, that means I'm off the hook." Sorry, but the truth is this: if you elect not to act, \textit{you are acting against the moral imperative to engage}. 
    \begin{quote}
        If you do not experience systemic racism, you are likely benefiting from it, whether by being more generously supported by your institutions, being assumed to belong, or, importantly, not bearing the significant mental and emotional burden of being subjected to racism.
        \vspace{-0.1in}
        \flushright{Gosztyla et.al. (2021) \cite{Gosztyla:2021}}
    \end{quote}
    
    \begin{quote} 
    If you are neutral in situations of injustice, you have chosen the side of the oppressor. If an elephant has its foot on the tail of a mouse and you say that you are neutral, the mouse will not appreciate your neutrality.
    \vspace{-0.1in}
    \flushright{Desmund Tutu (\textit{attributed, said before 1986)}}
    \end{quote}

    
    \paragraph{``We're all good people": Denial of the problem} \label{sssec:Pushback-DenialOfProblem} This argument usually takes the form of ``we know how to behave appropriately," or ``it's not bad \textit{here}," or ``I've never heard of this happening," all with the end argument of ``why is a code of conduct even necessary?" Lack of personal experience with harassment and discrimination does not preclude such behavior from existing. Study \cite{nasem2018} after study \cite{clancy-JGeophysResPlanets-Doublejeopardy-2017} after study\cite{aycock2019,ford:2011,ford:2019,sue:2008,griffin:2011, pewresearchcenter:2018} have demonstrated the stark disparities in experiences of minoritized populations within STEM, including physics specifically. This problem is not unique to nor excludes HEPA spaces. 
    
    \textbf{To be blunt - if you think that misconduct is not happening in your collaborations and communities, it is solely because no one has talked to you about it, and not because your specific community is exempt. 
    }

    
    \paragraph{``We already have rules against this!": Deferral to existing standards} \label{sssec:Pushback-ExistingStandards} ``The United States has laws in place to protect against workplace discrimination and harassment. Why are these rules not enough?" Yes, the U.S. has had laws in place to protect against workplace discrimination and harassment. The EEOC was created just after the passage of the Civil Rights Act of 1964, nearly 60 years ago, and yet minoritized populations are \textit{still} discriminated against and harassed in STEM. Beyond that, we have discussed above how collaborations often work outside of the traditional ``workplace": at collaboration meetings, conferences, workshops, summer schools --- all places which have, at best, limited protections through anti-discrimination ordinances. 
    
    Not only that, but we should expect our communities to do more than the bare minimum legally required. We should expect our communities to do more than the bare minimum to protect the underrepresented, the minoritized, the vulnerable among its members. We should expect our communities to prioritize safety and well-being \textit{even over scientific output}. There too many loopholes in workplace harassment and discrimination laws, especially as seen in our discussion of institutional policies (section \ref{sec:Institutions}).
    
    One caveat is mandatory reporting --- there should be explicit discussions about responsibilities before someone agrees to be responsible for reports. They should share what types of incidents they are required to report (for example, does it need to involve their specific institution), as well as who they must report to. This should be viewed as a potential conflict of interest, especially if the collaboration intends to keep reports confidential.

    \paragraph{``Wokeism" and ``back in my day": Minimizing the Actual Harm} \label{sssec:Pushback-MinimizingActualHarm}
    The term ``woke" became widely used in it's current form in the 20th century, meaning to be politically or culturally well-informed or aware. It originated in and continues to have strong ties to Black communities in the United States, especially related to anti-Black police violence. Aja Romano (Vox) writes that it ``has evolved into a single-word summation of leftist political ideology, centered on social justice politics and critical race theory"\cite{romano-Vox-Howbeing-2020}. More recently it has become co-opted by American conservatives who associate it with performative activism and with the right-wing theories of ``cancel culture" and ``political correctness." Indeed, if this is a push-back to a code of conduct, it says more about the speaker than of the code: it is another way of saying ``I am so attached to the systems that benefit me, I am so attached to my privilege, that I would prefer to devalue the comfort and safety of my colleagues rather than evaluate and change my own behavior." 
    
    This also includes discussions of ``they didn't mean it like that" or ``this is way overblown" or ``I wouldn't have been bothered by it," which usually revolve around the intent of the harasser. If this is common in your discussions, we encourage you to review work by Kate Manne on the topic of ``himpathy" or ``the excessive sympathy shown toward male perpetrators of sexual violence" \cite{manneGirlLogicMisogyny2018}. Questions like ``should the perpetrator have known better?" or ``did the perpetrator realize they were harming their colleague?" or ``what harmful experiences led the perpetrator to act this way?" decentralize the action and impact of that action\cite{Aurora:2019}. The focus is placed incorrectly on the reputation or emotions of the perpetrator, rather than where it should be --- on the feelings of the target and the rest of the community. The far more important questions to be asking are 
    \begin{itemize}\setlength\itemsep{0em}
        \item What impact does this behavior have on the target? On their work environment? On their job prospects?
        \item Is the community vulnerable to this type of behavior? What can be done to prevent this harm?
        \item What harmful experiences did this act create for people within and without the community?
    \end{itemize}
    Harm to a collaboration affiliate is harm to the collaboration. Again,  electing not to act \textit{is acting against the moral imperative to engage}, and choosing to remain neutral favors those engaging in misconduct.

    \paragraph{``We're changing already, isn't that enough?": Denying responsibility for activism}  \label{sssec:Pushback-DenyingResponsibilityForAcivism} This pushback is often categorized in the same vein as ``these changes take time". We heard during the writing of this paper sentiments that echoed ``the only people opposed to DEI are old and if we just wait, generational change will fix everything automatically" --- which is incorrect on multiple levels, least of all the fallacy that newer generations do not espouse harmful ideas.
    
    We acknowledge that there has been change. We discussed in sections \ref{sec:Introduction} and \ref{sec:Collaborations} that there are efforts ongoing to affect the culture and climate of HEPA spaces. But this change is not nearly enough, and it is not happening nearly quickly enough. We call on all academics who identify with at least one majoritized identity (that is: white, able, male, cisgender, and heteronormative identities) to move past passive allyship and instead take up the mantle of ``accomplices" against racism, sexism, and other forms of harassment and discrimination\cite{harden-mooreMovingAllyAccomplice2019}.


\subsection{Handling Misconduct} \label{sec:CollabHandlingMisconduct}
Disciplinary actions taken should correspond to the severity and frequency of the harassment \cite{nasem2018}. In theory, a variety of possible actions may be taken for violations, including: collaboration leadership speaking to the perpetrator, expert-led mediation, temporary or permanent removal from in-person meetings, temporary or permanent removal from authorship lists, or removal from the collaboration; many of these potential disciplinary actions are outlined in various collaboration codes of conduct. However, in practice, due to many of these actions are not possible. Collaborations do not have trained mediators on hand, nor the money to hire one. One might look within the collaboration for a mediator; however, the tight-knit nature of collaborations make navigating such relationships difficult, if not impossible, resulting in biased mediation. Additionally, physicists are not typically trained as mediators, which likely leads to negative outcomes for the person reporting misconduct.

Other, more serious disciplinary action that might have consequences for a collaboration affiliates’s career require an investigation, lest the collaboration governing body who implements these disciplinary actions be at risk for legal action (see subsection \ref{sec:legality}). However, collaborations cannot perform their own investigations into violations of codes of conduct as they do not have trained investigators, and physicists are in no position to be investigating their peers, especially if the misconduct was egregious.

National labs hosting our experiments may be the only legal entity that collaboration affiliates have in common, and one might hope that these labs would investigate violations of codes of conduct.  One might think to outsource the investigation to a lab or other institution; however, this idea quickly becomes complicated as discussed in section \ref{sec:Institutions}. Policies are not consistent between all labs, and investigations of harassment should not depend on the generosity of a particular institution. A victim should not have to hope that a lab has a policy to investigate whether or not either the victim or perpetrator must be employees of the lab, whether or not the activity was related to a hosted collaboration, whether or not the institution only investigates events that occur on institution campus grounds.

An independent third party investigator could provide a report to collaboration leadership, who could then vote on disciplinary measures. However, collaborations do not have funding allocated for this. It may be possible to request money from funding agencies, but as it would be requested on an as-needed basis, collaborations would need to wait until the money is available before hiring an investigator, leaving the reporter to fend for themselves until the investigation is completed. In addition, someone from the collaboration will need to seek out an investigator; there are no investigators on retainer at institutions or labs.

In short, guidance is needed at the federal, the institution, and the collaboration level as to the implementation of codes of conduct at the collaboration level, such that victims of misconduct are not falsely led to believe that these codes of conduct will actually protect them in the event that they seek action from collaboration leadership. In parallel, collaboration leadership should be up front with reporters about what the collaboration can and cannot do, so that the reporter's trust is not betrayed by a system designed to fail.

\subsection{Communicating Findings, and a Comment on Legality}
\label{sec:legality}

\recommendation{F1.3}{For collaborations and other institutions that are not beholden to a governing body in a legal sense, funding agencies should provide formal recommendations regarding enforcement of codes of conduct, including handling community threats, removal of collaboration affiliates, and leadership rights and responsibilities, and protections against legal liability for leadership that is responsible for that enforcement.}

As collaborations reside in a legal grey area, it may be easy to write off the liability of a collaboration when legal questions arise. For example, if an affiliate is sanctioned by the collaboration publicly, can the messenger be sued for defamation? Are those who imposed the sanctions open to legal liability, especially in the case where such sanctions may impact the financial or physical security of one or more affiliates? 

To be clear, although we cannot publicly identify these individuals, during the development of this work we did hear from senior members of the community who have sought legal advice regarding whether they are able to enact the disciplinary measures that are included in their collaboration's code of conduct without opening themselves up to a lawsuit --- \textit{the resounding legal advice was ``no"}. We encourage HEPA collaboration leadership to consider this when drafting and implementing a code of conduct, and \textbf{we pose the need for funding agencies to clarify the legal standing of collaborations when it comes to addressing misconduct, or at the very least provide a way to hold bad actors accountable without exposing collaboration leadership to legal action, regardless of the legal standing of the collaboration as a whole.}

Additionally, as discussed in section \ref{sssec:Pitfall-NotLegalEntities}, some legal advisors may suggest that collaboration leadership not speak publicly about the removal of a collaboration affiliate for fear of legal reprisal. This means, that even in the cases of particularly egregious behavior, or where members of multiple collaborations are affected, \textit{these findings cannot be communicated between collaborations, or even provided to the institutions of those involved in the report}. Without the ability to communicate, we cannot prevent cases of ``pass the harasser" between institutions, in which someone who violates the code of conduct in one institution and is removed joins another institution which has no knowledge of their previous behavior and sanctions \cite{flahertyReferenceChecksAhead2019,mervisNSFUnwittinglyHired}.
Even if a collaboration removes a member for egregious behavior, that person need not join another collaboration to be successful. They can still be invited to give talks, receive grants from funding agencies, and remain in the field to continue their egregious behavior with a different population of victims --- this is especially the case when collaborations have acted independently with no ability to inform the public of the incident(s), leaving whisper networks the only method of protecting the community. The only way to remove harmful members of the community \textit{from} the community, is for a centralized system to effectuate their removal. 

Less likely to be discussed is the legal argument \textit{for} collaborations to take swift action: if a collaboration knew about someone with a history of harassment or discrimination, and did nothing, does that open collaboration leadership to legal liability when that person repeats the offense? Victims of harassment have time and time again felt the need to go public with allegations or sue the offices and institutions that betrayed them (e.g. \cite{martinTitleIXInaction2020}). Most recently, three Harvard students sued the university for a ``decade-long failure to protect students from sexual abuse and career-ending retaliation"\cite{HarvardLawsuit22}, alleging that Harvard's investigatory procedures were used to enable the abuse of and facilitate blackmail by the accused professor. \textbf{It is naive to believe that collaborations who fail to protect their most vulnerable members would not open collaboration leadership up to litigation from victims. }

\begin{quote}
    When enforcing a code of conduct, we recommend people focus on protecting the safety of their community, not on retaining members of their community who are offenders. When offenders sincerely accept responsibility, want to repair the harm they did, want to prevent future harm, and prioritize community safety over their own desires and needs, that's a great outcome and it is reasonable to work with the offender. Otherwise, we recommend using deterrence and prevention: showing the community that you're serious about the code of conduct by ejecting anyone who seems likely to violate the code of conduct in the future, as judged by their actions and statements both within and outside the community.
    
    \hfill \textit{How to Respond to Code of Conduct Reports}, Aurora \cite{Aurora:2019}
\end{quote}

%% file: OmbudsInterviews.tex
As part of this work, we interviewed a set of collaboration ombudspersons or DEI representatives. We especially thank the following people for their contributions to this section: 
\begin{itemize}   \setlength\itemsep{.2pt}
    \item Ellen Bechtol, Outreach Specialist at Wisconsin IceCube Particle Astrophysics Center
    \item L. J. Kaufman, nEXO Ombudsperson
    \item Mark D. Messier, NOvA Equity, Diversity, and Inclusion Chair
    \item Joel Sander, Ombudsperson for SuperCDMS
\end{itemize}
        

\subsubsection{LZ collaboration}\label{sec:Implementation_LZ}
    In 2018, the LZ collaboration joined a growing number of other science collaborations and formally instituted an Equity and Inclusion (E\&I)\footnote{Collaboration working groups use different naming conventions, including E\&I, DE\&I, DEI, EDI, JEDI, etc. We maintain the naming convention as used by each collaboration to avoid confusion when referencing the cited Codes of Conduct.} Committee, composed of professors, postdocs, and graduate students, to proactively discuss and act upon ways to support a more equitable collaboration. While the committee is not expert on these topics, several programs have been implemented that align with advice and best practices from experts in the field. 

    Two collaboration-wide ombudspersons installed by the E\&I committee provide resources to any collaboration member needing guidance on conflict resolution or harassment. The ombudsperson system has benefited LZ in a number of ways. All collaboration members can, in confidence, report problems or seek guidance from designated senior collaboration members, with the goal of resolving issues early before they become more serious. Having two ombudspersons gives collaboration members an option of whom to contact, provides for overlapping terms which enable a smooth hand-off, and also provides for both a US- and a Europe-based ombudsperson to account for differing legal standards and rules. The overlapping terms also help maintain institutional memory of past issues and how they were (or weren’t) resolved. LZ is not the only collaboration with a system like this, and we believe an informal network of ombudspersons within the physics community would enable us all to learn from one another, identify common problems in the physics culture, and consider how to address unique challenges that arise in large international collaborations, especially those with significant field work.
    
    The LZ E\&I committee wrote a Code of Conduct, which was adopted by the full LZ collaboration, that details the expectations for all community members when communicating with collaborators or attending LZ events. Since the nature of the experiment requires collaboration members of many different career stages to work and live together at the experimental site at the Sanford Underground Research Facility (SURF) 
    in South Dakota, the E\&I committee has also helped to establish house rules that ensure welcoming and equitable behavior in the collaboration-managed living spaces and have implemented periodic check-ins with those on site using anonymous surveys also help to identify any issues before they become more problematic. 
    
    Self and group education on sources of and ways to address inequity is critical. LZ has an external speaker series during our bi-annual collaboration meetings, featuring speakers who are well trained in EDI topics, many of whom also engage in EDI-related scientific research. Past talks have focused on effective diversity programs\cite{dobbinWhatMakesDiversity2018}, understanding implicit bias\cite{shaumanImplicitBiasIts2019}, understanding and working with the Lakota community near SURF\cite{decoryUnciMakaTells2019}, effective outreach programs to underprivileged community members, and identifying and addressing current and historical inequity in undergraduate education by analyzing student educational achievement\cite{mckaySEISMICEquityInclusion2020}. To support the Strike for Black Lives \cite{ParticlesForJusticeWebsite,ShutDownSTEMWebsite,VanguardSTEM}, members of the E\&I committee facilitated a structured discussion on systemic racism, using the APS TEAM-UP report\cite{TEAMUPreport2020} as a guide. 



\subsubsection{IceCube collaboration}\label{sec:Implementation_IceCube}
The IceCube collaboration finalized their Code of Conduct in 2018. Unlike other collaborations mentioned in this text, IceCube brought in a Community Engagement Fellow from what was then under the AAAS and is now the Center for Scientific Collaboration and Community Engagement (CSCCE)\cite{CenterScientificCollaboration} to assist their DEI committee and create the Code of Conduct. This person was partially funded through the fellowship to specifically engage with the IceCube collaboration, and to drive change. Simultaneously, an ombuds program (Appendix G in the IceCube Collaboration Governance Document\cite{IceCubeCollaborationGovernance2022}) was developed and launched in 2019; and a Policy for Formal Complaints (Appendix H in the IceCube Collaboration Governance Document) was developed around the same time. The two ombudsperson positions serve as a resource for collaborators and hold all conversations in confidence. Ombuds can simply listen, offer advice, point individuals to resources, play a role in informal resolution of issues, or confidentially raise an issue with collaboration leadership. 

The IceCube Code of Conduct applies to any interaction between IceCube collaborators, whether virtual or in person, in meetings or conferences. All IceCube collaborations and workshops include the IceCube Code as part of registration, and attendees must acknowledge their agreement of the code to sign up for the event. 
 
IceCube has also implemented a separate EDI working group, currently co-convened by one dedicated EDI collaborator which has been successfully implementing events and policies since the code was created. Events have included: meeting meetups for women / gender minorities \& the LGBTQ+ally community, EDI-centered social media campaigns\cite{IceCubeInternationalWomensDayTweet2022}, availability of pronoun stickers at meetings, and workshops on career and science communication. They are also excited as they look to the future with potential mentoring programs for early career scientists as well as dependent care grants for collaboration meetings. Finally, IceCube is a founding member of the Multimessenger Diversity Network\cite{MultimessengerDiversityNetwork}; this organization brings together EDI working groups from a range of multimessenger astronomy collaborations in order to share best practices, provide resources, and engage with experts in EDI. MDN also submitted a contributed paper to Astro2020\cite{networkAstro2020APCWhite2019} whose recommendations closely align with the themes of this work.

\subsubsection{nEXO collaboration}\label{sec:Implementation_nEXO}
The nEXO Code of Conduct is available on their public-facing website\cite{nexocollaborationDiversityEquityInclusion}. It was developed over the course of six to nine months, and was approved by the collaboration board in 2020. The development team consisted of a set of motivated mid-career scientists who referred to existing codes, especially that of LZ (Section \ref{sec:Implementation_LZ}), during development. It also included members who had written codes or been a part of EDI initiatives in other HEPA collaborations. The nEXO Code of Conduct is now a standalone document, which in the future will be included in the formal nEXO onboarding process (currently, PIs are asked to share the Code with their trainees during local onboarding).

nEXO wrote two ombudsperson positions into their code, specifically such that they would be collaboration representatives, nominated and elected by the collaboration as a whole. Staggered terms allowed for the continuity of information, and nEXO ombudspersons received formal training in confidentiality and ethics from the International Ombudsman Association\cite{InternationalOmbudsAssociation}. The first goal of the ombuds is to determine if reports can be resolved by local resources such as institution offices. Systemic issues or patterns can be reported to the nEXO Code of Conduct committee and spokesperson. Reports to ombudspersons are considered confidential; even in the case where a reporter elects to escalate the issue to the spokesperson, the two ombuds maintain anonymity to the extent possible — even between each other. Mandatory reporting was also considered when developing the ombudspersons roles, namely that the nEXO Ombudsperson Guide\cite{nexocollaborationNEXOOmbudspersonGuide} specifically calls out situations where confidentiality is waived. In order to ``maintain institutional memory,'' ombuds keep a confidential log; identities are still only shared between ombuds with the explicit permission of the reporter. 

nEXO also supports a separate, standing DEI committee, which has instituted an undergraduate mentorship program, hosted guest speakers at collaboration meetings, and is in the process of developing a climate survey for the collaboration; nEXO plans to fund this survey through their host lab, LLNL. The committee also compiled a collaboration-wide information hub for DEI related resources, which has expanded to host information on job postings and other recruitment opportunities.

\subsubsection{NOvA collaboration}\label{sec:Implementation_NOvA}
NOvA is hosted by Fermi National Accelerator Laboratory (Fermilab). NOvA’s Code of Conduct\cite{thenovacollaborationPublicdoc11v1CodeConduct} was driven in part by the Fermilab Neutrino Division\cite{NeutrinoDivisionFermilab}. Development took about a year, and the final draft was voted on by the Institution Board (IB) in Fall 2019. The development team was intentionally assembled with a diversity of gender, region, and career stage. The code was strongly based on literature and previous codes including DESC\cite{LSSTDarkEnergyCodeOfConduct} and APS\cite{CodeConductAPS}, and was circulated for collaboration comments before submission to the IB for voting. Parts of the code were also reviewed by local Title IX offices and Fermilab Human Resources. The code is publicly available on the NOvA homepage, is included in onboarding emails to new members, and is reviewed at each collaboration meeting. New conveners are reminded of the responsibility they have in maintaining a welcoming and inclusive climate.

The reporting system was developed with the intention of providing support for those collaboration affiliates who report and giving them control over what would happen with the report. Potential candidates for the EDI chair position are nominated by the collaboration, but appointed by the NOvA spokespeople; their role is primarily advocacy, such that reporters feel supported throughout the process. The appointments require endorsement from early career members, who could be particularly hesitant to escalate reports to senior affiliates. The positions are two-year appointments, staggered, with the option for renewal. Reporters are encouraged to report violations which may require investigations to Fermilab HR to be continued. 

The EDI chairs also have more formal responsibilities to the collaboration. Chairs meet with spokespersons to advise on meetings and plenary sessions, and convenership appointments. The chairs also report to the IB with statistical information about reports they have received, based on the categories identified in the code of conduct. Finally, chair reports to the collaboration meeting have become a “fixture of [the meeting] structure," which also include discussion of code of conduct related topics.

\subsubsection{SuperCDMS collaboration}\label{sec:Implementation_SuperCDMS}
The code of SuperCDMS is publicly available\cite{SuperCDMSCodeConduct2021}, and includes broad categories of violations as well as the implementation of an Ombudsperson within the SuperCDMS collaboration, whose role is ``to facilitate informal and unofficial resolutions, as well as the role of the Collaboration and its mechanisms for official procedures and actions when violations are alleged.'' The ombudsperson and their appointed ``safe persons'' serve several roles. First, they are available to provide informal solutions when a person coming forward prefers an informal solution. Second, they are victims' advocates during the resolution process, which only begins when a concern is brought to the Spokesperson or member of the Executive Committee (EC). The EC gathers confidential information and may include interviews with affiliates involved in the report. The EC then determines if a violation occurred and will seek to rectify the situation through structural or individual solutions. Anonymity of the violator may be revoked at this stage so that the EC may provide a short summary to the collaboration of violations and actions taken, including letters of censure or expulsion from the Council, Board, or Collaboration. Records are also kept within the EC to track patterns of behavior. Third, they act as a ``preventer'' by doing background ``culture setting work'' to try to prevent this type of occurrence from even happening, by providing people a friendly (and confidential) ear also on ``small things'' that may or may not evolve into something ``big'' and by providing information and by bringing ``good practice and appropriate conduct'' to the attention of everyone.

The code of conduct was developed over several months in 2019, and has been updated in 2021 to address gaps identified during collaboration review. Notably, the code of conduct applies to ``all collaboration interactions, including meetings, social events, and one-on-one interactions as well as the use of mailing lists, forums and social media'' \cite{SuperCDMSCodeConduct2021} . Additionally, an annual poll has been developed to receive information on how the collaboration was functioning — this was intentionally implemented as part of the code of conduct development process — including experiences within the collaboration, awareness of resources, and level of comfort reporting a violation. There is also a strong emphasis on how the collaboration is functioning at the affiliate level; a working group task force was asked to evaluate how the work of the collaboration is done— are working groups functioning, are affiliates receiving appropriate and timely feedback, is the culture working for junior members? The survey was developed with \textit{heavy reliance on an expert in organizational psychology}, who was better able to provide feedback on how to probe the desired themes and helped the collaboration drive organizational change; \textit{a funding agency could support this expertise for other collaborations, which would save time, money, and energy and produce knowledge which could then be implemented field-wide}. 

%% file: EngineeringRoles.tex

The communities that build and maintain the accelerators and detectors are usually associated in physics laboratories with instrumentalists that, in turn, have a considerable kinship with engineers. These communities themselves exhibit a complex structure and are divided in subcategories.  Instrumentalists can be considered a subset of detector and accelerator specialists, and the  watershed line between their research expertise and that of experimentalists (and, sometimes, theorists) is rather blurred. Whereas the majority of instrumentalists (and a certain fraction of pure engineers) are holders of advanced degrees, they are nevertheless considered  non-scientists and are not normally allowed to contribute to running physics experiments, analyzing data or developing high-level theory\cite{pronskikhBlurredEngineeringIdentities2021}.

\subsection{Disparities between scientists and non-scientists}
There  are distinctions  between experimental scientists and ``non-scientists” that manifest themselves throughout the course of a project. At the stage of creating experimental setups in physics, this becomes evident in the attention given to the design of the experiment. Although they both study processes in the detector on computational models, experimentalists focus more on aspects related to future searches for a useful signal (e.g. reconstruction of events occurring in the detector by triggering numerous sensors), while ``non-scientists” focus on aspects related to ensuring the overall operability of the installation. The second difference arises during operation of the setup when measurements are being made. The scientists participate in the data acquisition and subsequently process and analyze the data while the instrumentalists begin to focus on the creation of other installations and instruments. Nonetheless, both scientists and
the so-called ``non-scientists" in experiments are involved in
similar technical research activities, which makes the
distinction between these groups blurred and often formal.
Taking into consideration the fact that despite the similar nature of research work, the hierarchical subordination of the second group of the first, as well as the deprivation of instrumentalists and other ``non-scientists" of the right to independent ``non-directed" search and the social status of a ``scientist" in the community, coupled with a lack of mobility and exclusion
from important kinds of research activities, creates a situation of injustice similar to those discussed in this chapter with situations of gender and racial injustice and inequality. Also, it creates a non-productive tension in the community, associated with unwarranted contempt and disdain from the groups that are formally attributed to ``scientists" to ``non-scientists."

The basis of the external distinction between engineers and scientists in a research laboratory is rooted in the orientation of their constructive activities.  Engineers focus on activities that are artificial and technical in nature while scientists focus on those more closely tied to the natural phenomena under investigation. At the same time, the engineering nature of scientific labor, that which involves constructing and commissioning an apparatus, turns out to be characteristic of both  scientists and engineers. Therefore, it is only the formal orientation of the activity toward artificial aspects and functional roles which, as a rule, serves as the basis for the refusal of the community to allow engineering specialists to more fully participate in experiments and analyze the data collected by them. This exclusion of engineers and other formally non scientist specializations in megascience from the most valuable roles and practices can be considered a “participatory injustice" \cite{pronskikhBlurredEngineeringIdentities2021, hookway_2010}.

As a mitigation of such participatory injustice, an approach has been suggested to overcome participatory injustice by creating joint projects for non-scientific and scientific specializations in which they cast themselves in equivalent roles. Two specific directions for the mitigation of inequity to knowledge production practices can be explored during the course of Snowmass.

\paragraph{Imitation Games} The first direction is an implementation of Imitation Games (IG) for scientists and engineers\cite{collinsImitationGameNature2017}. IG implies roles of “imitator,” “honest participant,” and “evaluator”: all participants are presented with questions regarding instrumentation, running experiments, data analysis, and physics theories. “Honest participants”, belonging to a certain community, answer the questions in a straightforward manner relying on their own professional expertise and gut feeling. “Imitators” attempt to answer the questions as if they belonged to another community in a manner indistinguishable from a member of that community. For example, an experimentalist “imitator” pretends they are an engineer and answers how they think an engineer would answer. “Evaluators” then try to distinguish whether the answers were fake or not and explain why they considered answers true or fake. IG can be played both in person and online and is a proven method of facilitating mutual understanding between different social groups and overcoming prejudices. As such, during the course of Snowmass, this can be developed and carried out as a community intervention both during virtual town hall style meetings and at the larger in-person gatherings.

As example, we suggest the following kinds of questions for IG:

\vspace{0.1in}

\begin{tabular}{p{0.2\textwidth} p{0.7\textwidth}}
    For everyone &  Have the experiments at LHC helped identify new particles with the properties inconsistent with the Standard Model?\\
    
    For non-scientists & Do you envision any significant
    academic or experiential deficiencies that would disallow you to take over the functions performed by scientists?\\
    
    For everyone & Please, give examples of the qualities possessed by scientists and at the same time absent or lacking in engineers?\\
\end{tabular}

\paragraph{Systematic Change} The second direction is that institutional barriers between communities can be lowered, and vertical mobility in the scientific community facilitated. Currently, in the US national laboratories the mobility between those on the scientist and the non-scientist tracks is almost non-existent. For example, it is incredibly challenging for an individual even with a formal scientific background but on a non-scientific track to pursue a career in a scientific direction (like data analysis or theoretical studies). Presence of rigid borders between subcommunities accompanied by sometimes arbitrary or willful criteria of ascribing researchers to either scientists or non-scientists creates participatory injustice. This needs to be alleviated in order for the entire field in the US to flourish.


\subsection{Science is not done in a vacuum}


The people who participate in and contribute to the success of our scientific endeavors are individual and human beings, who besides their academic curiosity and professional pride want recognition, respect, compensation, to feel they belong to a community and that their contributions matter.

In addition to the academic and scientific aspect of collaborative science, there is a vitally important social, mentoring, and networking aspect. This is important at all stages, but particularly evident for those early in their careers. Young technicians, engineers, operators, and physicists will experience very different levels of support, inclusion, mentorship, recognition, professional development, and opportunities of all kinds over their careers, and this divide continues and its cumulative effects multiply over the years. The pandemic has only increased this isolation and divide for many. ``Essential" technical and operational staff often found themselves vitally needed and completely forgotten at the same time.

We must invest in the members of our community. In the same way that systemic inequalities in matters of gender, race, or other minoritized identities add stress and stand in the way of better science and happier people, these distinctions in professional respect, inclusion, and appreciation inhibit communication, development, and the free flow of ideas. There are times when certain ``scientist"/``non-scientist" distinctions are important and useful; but there are other times when this artificial caste system mainly serves to hurt communication, productivity, and morale. All contributions are necessary to the success of the enterprise, but often, only some receive recognition. All careers should be supported and developed, but often, some are left to stagnate. Without support, skilled professionals may depart to other environments in which they feel better appreciated or compensated. This is a loss both to the field that could have benefited from their contribution and to the individuals who were not able to stay in a field about which they may have been as passionate and innovative as anyone.

\subsection{Why We're Talking About This Now: Disparities Beyond Scientists and Non-Scientists} 

\begin{center}\textbf{\textit{Institutionally-designed hierarchies exist within HEPA spaces.}}\end{center}

Beyond the discussion above, the hierarchy that probably comes to mind first is the aspect of seniority:  undergraduate student, graduate student, postdoctoral researcher, followed by adjunct faculty, tenure-track faculty, and tenured faculty for academia, and junior and senior research staff for laboratories. This hierarchy has developed over time with the assumption that experience breeds knowledge and wisdom which should be passed down the ladder to those who are just beginning --- but it also hinders the flow of knowledge the other way, especially on topics related to diversity (such as that discussed in section  \ref{sssec:Pushback-DenyingResponsibilityForAcivism}). 

This hierarchy also prefers the identity groups that make up the higher rungs of the ladder. In a simple sense, this may take the form of advice being given to graduate students based on decades-old experiences, or experiences that were derived from different lived experiences. More insidiously, the higher rungs actively (consciously or unconsciously) prefer to facilitate the advancement of 
junior affiliates who share their identity groups --- as evidenced by the failure to retain underrepresented minorities through the academic `pipeline' \cite{rieglecrumb-EducationalResearcher-DoesSTEM-2019}, the evidence for bias in the interview (e.g. \cite{dutt-NatureGeosci-Genderdifferences-2016, mossracusin-ProcNatlAcadSci-Sciencefaculty-2012}), grant funding (e.g. \cite{rissler-BioScience-GenderDifferences-2020}), and publication processes (e.g. \cite{caplar-NatAstron-Quantitativeevaluation-2017}), persistent cultures of overt and covert discrimination (e.g. \cite{davis-blackintheivory, Rosa-thesis-2013,chambers-DifferentKind-2017}) and lack of intential mentoring\cite{monroeGenderEqualityIvory2014}. This results in a field which prefers itself to be white, male, able, cisgender, and heteronormative.

The hierarchy of scientists and non-scientists may be more familiar to the reader, but it also can provide insight into the virtual hierarchies and inaccessibility of other identity groups. It goes further --- the intersection of identity groups which are marginalized within even these hierarchies continually seeks to rank us according to this imagined hierarchy of ``pure scientific contribution to the field." HEPA needs to rid itself of this ``meritocratical" viewpoint, not only because of its benefits to the field, but because it is right and it is long overdue. 

\begin{quote}
    What we really do not need in STEM is more of the same type of students from the same institutions, taught by the same professors, learning the same curriculum, working at STEM institutions where everybody looks (and quite possibly thinks) similarly. As long as there is widespread reluctance in these fields to address the insidious, complex effects of structural racism, which range from the individual to the institutional, STEM education will ultimately result in less robust, innovative, creative STEM industries and outputs.
    
    \hfill McGee (2020) \cite{mcgeeInterrogatingStructuralRacism2020}
\end{quote}


%% file: HEPsoftware.tex
The Snowmass 2021 effort provides an opportunity to critically examine the data analysis practices and culture in our community. While there is little doubt that past efforts have led to some truly amazing scientific discoveries, we suggest that practices could be improved, particularly in analysis software and the culture around the use of that software, that leads to more productive and efficient workflows while also creating a more inclusive environment. Snowmass offers us an opportunity for us to re-think how we approach our software  and how we design the training and onboarding for new colleagues.

Big Science often involves experiments that take many years, even decades, to plan and run. These large, complex experiments result in large, complex software stacks that  have often been designed by a committee of hundreds, if not thousands. Code is usually written by a core group of experts, who have a wealth of knowledge and expertise. This knowledge  filters down to new collaboration affiliates whose nearest-neighbor experts (usually senior students and post-docs in their own group) do not always share this knowledge. The experiments have a long timeline and so their  software stacks will span shifts and improvements in software and hardware and will incorporate those changes in {\it real time}. These improvements save time and increase experimental sensitivity. However, these improvements are often rolled out in a way that make it even more difficult for students to adapt to.

Because documentation is often the last thing to be written and because most grad students and post-docs who write the documentation are usually not trained as teachers or mentors, the documentation does not always serve the needs of the boots-on-the ground analysts who can be distributed all over the world. This results in graduate students spending the first 2-3 years of their PhD simply learning an evolving and changing framework just to access the data and MC in their experiment. This is the norm in most areas of HEP, and so PIs and others tend to accept this environment as simply an immutable fact of nature/the system.

This environment can be particularly frustrating for women, students of color, and other under-represented groups. The barrier to entry for a new researcher trying to understand and contribute to such a complex stack is often viewed as a rite of passage, something that everyone has to go through. Newcomers are given complex code to copy and edit, pointed to confusing or non-existent documentation (often in the form of commented code), and told to ``go ask someone" how to proceed. This is a system that exacerbates imposter syndrome - the feeling that you are not as good as everyone around you. If a young researcher doesn't understand what their code is doing, how can they consider themselves to be a credible scientist? The way we structure software development in our collaborations favors personality types that are comfortable asking for help, often at risk of being patronised in very public forums (e.g. Slack, Hypernews) or those who have the confidence to barrel through without really understanding what they're doing.

While there can be a real sense of achievement in understanding even a corner of the software used in a large collaboration like LSST, DUNE, or any of the LHC experiments, too often the experience leaves a researcher demoralised and disillusioned about the process of science. This stress can be mitigated by an involved and supportive mentor or PI, but not every group has that. Promising scientists are leaving the field as a direct result of the software culture, and anecdotal evidence suggests that this is more likely to happen to members of underrepresented groups in HEP. We are personally aware of students, mostly women, who left the field after their PhD (or even before completion, opting to take a Masters instead) in part because of their experience with software in their experiments. And even students who push through, express frustration at their experiences, especially when they are told by more senior members of their collaboration that ``this is just how things are".

The authors who proposed this work (Bard, Bellis) were motivated by their own experiences and stories from colleagues. 
We planned on conducting a more thorough survey of current HEP practitioners as well as those who have transitioned to
non-academic careers or non-HEP academic careers. However, the demands of work and family in the pandemic environment 
limited our efforts. As such, we decided to limit the scope of this section, hoping that it can still be a sign-post for future work.

First, we want to acknowledge the efforts of the community to effect positive change in how new students are onboarded,
even going back to the BaBar Analysis School, in which new collaboration affiliates were led through a week of lectures and hands-on
activities using the analysis software. This type of workshop is carried on in many experiments today. The CMS Data
Analysis School (CMSDAS) runs in a similar fashion, culminating in small groups performing a ``full" analysis in a
roughly two-day period. And experiment-agnostic efforts like the HEP Software Foundation, is working to produce
tutorials on concepts and programming languages\footnote{\url{https://hepsoftwarefoundation.org/training/curriculum.html}}, inspired by the Software Carpentry group\footnote{\url{https://software-carpentry.org/}}.

But even in a week-long workshop, organizers need to decide what {\it not} to teach, and so participants are still provided
with pre-written examples. These are {\it incredibly} helpful and we are not denigrating the work of many people
(including one of the authors (Bellis) of this section). But they can only provide so much scaffolding and so 
collaboration affiliates still spend the bulk of their graduate student or post-doc (or even undergrad) time finding their own 
way through the experiment's software ecosystem. 

The challenges we have commented on above are not unique to HEP. While we were not able to do a deep dive into 
the scholarship in this field, we were pointed to two researchers who study software communities and we hope that 
there is opportunities to engage their work in the future. Denae Ford Robinson is a Senior Researcher at Microsoft Research 
in the SAINTes group and an Affiliate Assistant Professor in the Human Centered Design and Engineering Department at the University of Washington. From her bio\footnote{\url{https://denaeford.me/bio/shortbio}}: 
\begin{quote}
    ``Her research lies at the intersection of Human-Computer Interaction and Software Engineering. In her work she identifies and dismantles cognitive and social barriers by designing mechanisms to support software developer participation in online socio-technical ecosystems. She is best known for her research on just-in-time mentorship as a mode to empower welcoming engagement in collaborative Q\&A for online programming communities including open-source software and work to empower marginalized software developers in online communities." 
\end{quote}
Her publications
include ``Is My Mic On?" Preparing SE Students for Collaborative Remote Work and Hybrid Team Communication~\cite{9402185}
and ``Including Everyone, Everywhere: Understanding Opportunities and Challenges of Geographic Gender-Inclusion in OSS"~\cite{9466393}.

Amy J. Ko is a Professor at the University of Washington Information School and an Adjunct Professor at the Paul G. Allen School of Computer Science and Engineering. Topics that she engages with are {\it Justice-focused computing education} and 
{\it Broadening participation in computing}, which we expand upon by quoting from her website\footnote{\url{https://faculty.washington.edu/ajko/}}: 
\begin{quote}
    ``The world is full of structural barriers to developing interest in and learning computing. My lab investigates these barriers and ways to eradicate them, including issues around accessibility, mentorship, culture, identity, and resources, with the goal of increasing the diversity of people in computing disciplines."
\end{quote}
Her publication list is equally relevant with work on ``Explicit programming strategies"~\cite{LaToza2020}
and ``Individual, Team, Organization, and Market: Four Lenses of Productivity"~\cite{Ko2019}.

Even a minimal interaction with these two scientists' professional experience demonstrates that there are challenges
in coordinating and engaging a wide range of individuals into large software projects, whether open-source
or large companies or, as we well know, within scientific communities. As the HEP community moves forward, we hope that 
time will be found to perhaps reach out to these or similar researchers so that we can learn from their work. 

\subsection{Anecdotes from Former Community Members}

While we were not able to conduct a proper survey, we still wanted to share some experiences by former members
of our community. This is admittedly a biased sample, as we reached out to specific colleagues who we
knew had experienced some of these frustrations. That being said, we recognize that these
are real experiences of our colleagues and as such are still valid. 
Whether or not these anecdotes are representative of a significant
number of researchers must be left to future work. 

We provide feedback from two individuals who worked on HEP experiments for some of their time in graduate school. 
In an effort to allow them anonymity, we only mention that their experiences span the past 15 years and that
they both identify as either women or a different gender minority. We received their permission to relate their
story as presented here. 

\paragraph{Individual \#1}
\begin{quote}
\begin{itemize}
\item As an undergraduate researcher, I was fortunate to work in a lab where running existing code was discouraged. There was an awareness of the culture of HEP to have students work with existing code blocks and software and part of my education was to recreate these with my own knowledge and skills. When not possible, this allowed for more energy to dig into documentation and poorly commented code when necessary. My energy to approach CLUSTER/GRID SUBMISSION TOOLS for submitting jobs was high as I hadn’t been working with this documentation as my main job. 
\item When I got to graduate school, I was in a LHC EXPERIMENT lab that did not value this. In my first research experience, I attempted to write my own code to perform analysis and was chastised for taking longer than it should take using existing code. When I attempted to use the existing code, I spent all of my time doing research reading documentation, tracing function and variable definitions through nested code, and asking questions to other graduate students. This did not appeal to my desires for exploration, creativity, and fun when performing my research. I wasn’t interested in this work I was putting in and the work became draining. 
\item In my next project, I was running code written by students one and five years ahead of me in the program. No documentation was written to pair with these files. I was told to ask the students for guidance. I did not want to eat up all of their time with endless questions and tutorials, so I spent a lot of time debugging on my own. When we did meet, there were diagrams drawn and maps for the code but I still had questions that remained unanswered. The barriers to exploring the questions I had about my data limited my excitement for the work. 
\item This research group had other cultural issues that made the group environment difficult. Aggressive lines of questioning in group meetings made students defensive and often ended in arguments. This made approaching anyone for assistance mentally taxing. Asking another student, professor, or post doc for help was like preparing for battle. I wanted to be excited about the work I was doing and instead, it was survival that was driving my work. 
\item Finally, I chose to leave the LHC EXPERIMENT group as I no longer had any drive to continue. I was experiencing extreme burnout and doing any amount of work drove me to tears. The code block approach to learning was compounded by the environment I was in and I was no longer able to continue my relationship with LHC EXPERIMENT. At this point, I turned to A NON-COLLIDER EXPERIMENT group to continue research. This group was friendly and curious. Code was well documented and I wrote all of my own code once more. I had a friend in the group who actively made time to work with me when I was stuck. My questions were appreciated as a signal of my love of the work. The environment at NON-COLLIDER EXPERIMENT is extremely
different than THE LHC EXPERIMENT GROUP and it was very sad to say good bye to the experiment I started my HEP career in.
\end{itemize}
\end{quote}

\paragraph{Individual \#2}
\begin{quote}
\begin{itemize}
\item I got into physics because I loved math and learning a bit about cosmology and string theory during senior year of high school blew my mind. And when I discovered particle physics in college, I was captivated by its theoretical elegance, teamwork and collaboration, chances to travel for conferences and the experiment, and opportunities for doing physical work fixing and upgrading the detector.
\item I took only one programming class in college, the intro to C class, which I didn't do great in, and was told by pretty much every advisor I had that I didn't need any more, that I'd pick it up as I go.
\item That was terrible advice. Of course I'd pick something up, but the fundamental ideas of how to find and solve bugs, how to write clean and readable code (to preempt bugs), how to read others' code, how to think of software as a whole, all of that should have been a course. Instead I painfully kluged my way through grad school. Stack overflow was also not as big as it is now and I didn't get to actually collaborate with anyone until I think the last year of grad school, so I was fully alone and adrift for years.
\item I hated C++. I couldn't understand why we didn't have CS people properly writing common libraries that I could just use for my higher-level analysis. I even asked about that and was told that CS people wouldn't understand the physics we need. 
\item I hated having to ``write complicated, excruciatingly atomic step-by-step instructions for the world's stupidest toddler who will do exactly everything you tell it to do completely literally," as someone on reddit phrased it. I don't think like that. I don't. I hate breaking seemingly obvious things down step by step in an orderly fashion. I love inspiration and creativity and beauty.
\item Being handed a topic (I didn't even get to discuss choosing one!) and left fully alone for years, not knowing who to ask for help, being afraid to ask for help because I felt that would reveal how dumb I was and how much I didn't deserve to be there, was awful. Being told I wasn't allowed to travel to conferences, even to practice presenting, because I needed to focus on my analysis without understanding how exactly, was awful. Being told I wasn't allowed to physically work on the detector, again because I had to ``focus on my analysis", was demoralizing. 
\item Contrasting that experience with being a data scientist, well a few things were similar. Not being listened to when I asked for what I want, for example. Not having a code review when I know that would preempt my own and my colleagues' bugs. Not getting learning resources when I asked for them.
\item But working with Python, and in a team that needed to use my code, is a completely different beast. I had weekly or biweekly meetings where I described to the client the upgrades in the algorithm, and when I needed help I was assigned it. The team discussed appropriate libraries to pull in. If someone didn't know what to do, they'd ask and there'd be no penalty, someone would just show them. Mostly. And Python is much more like Matlab, i.e. more directly doing math, than C++ was. It's still not as pleasant as doing math directly, but the instructions are slightly less atomic and the form of instructing can be a little varied.
\end{itemize}
\end{quote}

While the details are different, a commonality is that both of them felt that the software was barrier to doing what they 
{\it wanted} to do, which was physics, and that they did not always know who to go for help. Again, we stress that 
the situation is improving in many experiments and in many groups, but the improvements are most likely not
universal. 

\subsection{Where We Go From Here}
We are not experts in this field and so our comments must be taken with a grain of salt. But we have some minimal recommendations
for either individual experiments or the community as a whole. 
\begin{itemize}
    \item \textbf{Engage with other researchers.} We have identified two scientists who study large communities who
    engage in software development, Ford and Ko, and would strongly encourage the community to reach out to them
    or similar researchers.
    \item \textbf{Consider hiring scientific writers to produce better documentation and training modules.} 
    This would be a very different approach
    to how things are normally done where everything is produced ``in-house". But there are people who learn how to do this
    and if freed of the demands of performing an analysis, could probably do this better than many of us. 
    \item \textbf{Regularly re-evaluate what parts of the analysis workflow should be provided as ``services" and what should be left to analysts.} For example, does it benefit every analyst to have to continually re-run their skims 
    as new corrections are performed? Or to modify their workflow as new submission tools are developed? With an
    eye toward a ``greener" compute model, should we work to minimize how often analysts have to go back upstream
    and instead provide more centrally-produced skims, made to-order, so that analysts can focus on 
    the physics and novel approaches to extracting results? While the shift from C/C++ to python has helped
    lower the threshold for many researchers, new developments in accelerators and decorators are adding an 
    increased level of complexity (though providing significant gain in speed) and analysts need time 
    to learn these new tools. That means that we might have to offload some of the work that almost
    {\it every} analyst performs in the early stages of their work
\end{itemize}

While there is no obvious solution to ensure that all collaboration affiliates can contribute equally to the software ecosystem and feel supported in those efforts, we are encouraged that more and more people are realizing this is an issue to be tackled. There are already many projects and workshops among the experiments that are trying to address these challenges and we hope this discussion and this section can be part of those efforts. 


%% file: NonCodeTopics.tex
\subsection{Expertise and Compensating Experts}\label{sec:ExpertsAndExpertise}
    
    We begin this section with what is likely the most important takeaway from this paper: \textbf{that physicists should not be doing the job of organizational psychologists and DEI experts}. Our expertise is in high energy particle physics and astrophysics, and not in the broad fields of sociology, history, or psychology. Our best (and truthfully, only) solution to addressing both inequities in HEPA as well as deficiencies within HEPA culture, relies on the advice and recommendations of paid experts. We also call on funding agencies to require such coordination as part of funding documentation: inclusion of climate-related topics into safety components of `Operational Readiness Reviews,'' ``Conceptual Design Reviews,'' or similar documentation.
    
    \vspace{0.1in}
    
    \recommendation{F3.1}{Funding should be made available to both engage with and compensate such experts. This can take the form of independent grants, but more effective would be the inclusion of climate-related topics into safety components of collaboration ``Operational Readiness Reviews,'' ``Conceptual Design Reviews,'' or similar documentation submitted to funding agencies.}
    
    \recommendation{C3.1}{ Experts should be adequately integrated into HEPA communities, including collaborations, such that their expertise can be applied effectively. This may take the form of an official collaboration role like a non-voting member of a collaboration council.}
    
    Putting this call aside for a moment, even within our existing culture, physicists who do this work are not given proper recognition or respect. Many institutes and collaborations have `Diversity, Equity, and Inclusion’ committees, yet the work done within these bodies tends to not be implemented and the individuals doing the work are not rewarded in a professional capacity. Additionally, the current culture of physics categorically places DEI work as an optional `service’ and sometimes as a `distraction’ from research rather than a necessary and important aspect of the job. In fact, there is evidence that these service activities fall heavily on the shoulders of underrepresented groups and negatively impact the promotion or advancement for the engaged individuals.
    
    Finally, we must acknowledge that performing diversity and inclusion initiatives without expertise can very well harm the populations they are trying to help. Expertise should be considered a mandatory part of policy development. 
    \begin{quote}
    Too often, people approach transformative justice as a neat idea they'd like to experiment with in their community, and end up harming the most marginalized members of their community instead of helping them. If you want to use transformative justice in your community, we recommend you only do so if you have several people who have formal training in transformative justice available to work the process, and you only attempt the process with members of your community who have a strong incentive to stay inside the community and to treat other members of the community well.
    \flushright{How to Respond to Code of Conduct Reports \cite{Aurora:2019}}
    \end{quote}


\subsection{Diversity in Leadership and Representation}\label{sec:DiversityInLeadership}
    HEPA communities should be asked how their leadership is designed, how affiliates are trained and selected for leadership. Who actually ends up in charge? What are the short- and long-term repercussions of this choice?   \footnote{The authors credit \cite{Gosztyla:2021}, which details "Responses to 10 common criticisms of anti-racism action in STEMM" for a substantial portion of this section's literature review.} 
    
    As discussed at length in section \ref{sec:EngineersAndScientists}, inherently hierarchical structures such as seniority, scientists / engineers (section \ref{sec:EngineersAndScientists}), racial-, and gender-hierarchies offer to maintain the myth of ``meritocracy" in HEPA spaces. These must be replaced with policies that advance equity over an ill-defined conception of ``merit." Scientific ``meritocracies" do not exist. Even the man who coined the term ``meritocracy" meant it as satire: specifically that a meritocracy would appear at face value to be equitable but would rather continue the cycle of disenfranchisement through systemic processes designed to deprive minoritized populations of education and, therefore, upward mobility \cite{young-TheGuardian-meritocracy-2001,Gosztyla:2021}. Instead, success depends strongly on external factors which include but are not limited to race, gender, socio-economic class, and disability; the stress of these external factors contribute to greater cognitive load which in BIPOC has been associated with illness and reduced productivity\cite{eaganjr-JHighEduc-StressingOut-2015, williams-JHealthSocBehav-StressMental-2018, zivony-NatHumBehav-Academianot-2019}, which further distances BIPOC researchers in terms of ``merit". 
    
    And within larger collaborations, affiliates can often be rewarded by ``internal promotion” which does not come with monetary benefits but where individuals are given increased responsibility and greater visibility as an expert on a certain topic. This, in turn, can be leveraged for professional advancement. This may include assignment into leadership of working groups, but also may include the assignment of influential tasks tasks (such as which graduate student receives the Nature-worthy thesis topic). The manner in which these decisions are made is in no way uniform and often-times relies heavily on individual professional networks, thereby not rewarding ``good science” but ``who knows who.” It also means that the recognition of junior colleagues is more directly reliant on the power dynamics with the senior colleagues most closely connected to their work. This is inequitable and prone to creating ``in” and ``out” groups, sometimes referred to as ``gatekeeping,” and can be done intentionally or unintentionally. Leadership drawn from the conception of ``merit" (for example, the number of publications generated or cited, or admission to a top university) fail to account for such factors and are inherently biased towards those affiliates whose race, gender, socio-economic status, or disability did not influence their productivity. These choices perpetuate inequality in leadership, which is a driving factor among hiring choices and additional opportunities, and ultimately the next pool of potential collaboration leaders and field-wide experts. 
    
    We call on the community to develop methodologies, in association with experts in organizational psychology and sociology, to better address hiring and leadership disparities in HEPA. Collaborations, especially those which host multiple separate working groups, should be required to perform climate evaluations to ensure that advancement within the collaboration is equitable and that interventions are taken where necessary to improve the advancement of collaboration affiliates. 
    
    \vspace{0.1in}
    
    \recommendation{C2.1}{Reviews of community climate should include an evaluation of how leadership is selected within HEPA collaborations (e.g. assignment of high-impact analyses \& theses topics, convenership of working groups, public-facing roles representing the collaboration such as spokespersons or analysis announcement seminars)}
    
    \recommendation{C2.2}{Reviews of community climate should include the valuation of sub-community contributions. This includes the participation of ``non-scientists" in community engagement and authorship, community perceptions of operations and service work, the development of onboarding and early-career networks, and implementation of policies toward equity in information sharing and software.}


\subsection{Onboarding / Mentoring Networks (Including Early Career ERG)}\label{sec:OnboardingAndMentoring}
    
    
    The culture of our community cannot be assessed without acknowledging the existing hierarchical structure and power dynamics in academia that all too often allow the detrimental behaviors to permeate without recourse. In many systems, graduate students and postdocs report only to their advisor who solely controls their funding. Others have instituted mentorship or co-advisors for graduate students and/or postdocs that offer additional support throughout the full length of the program. The effectiveness of these programs can be assessed through questions gauging how included students/postdocs feel, what their support structures look like, if their needs are being met, and if they feel like they could report a concern with their advisor to their department/institution without it negatively impacting their careers. Including such questions in a climate survey could result in best practices for institutions to maintain mentorship and retention of graduate students and postdocs while providing a method to circumvent destructive power dynamics. 
    
    This topic is discussed briefly in section \ref{sec:InfoSharingHEPSoftware} with regards to onboarding into HEPA collaborations, and we'd like to expand on this theme here. Formal mentorship roles within a collaborative environment would go far --- not only towards addressing discrepancies in career trajectories post-PhD \cite{mulveyPhysicsDoctoratesSkills} but also toward a sense of belonging within physics communities. There is extensive work done on effective mentoring practices in physics (e.g.\cite{singhInclusiveMentoringMindset2021}), that this work does not have time to review, but could be addressed as part of a collaboration services office (section \ref{sec:CollaborationServices}) or cross-collaboration network. 
    
    Early career resource groups have been implemented in HEPA collaborations, such as in ATLAS\footnote{\url{https://atlas.cern/authors/atlas-early-career-scientists-board}}. Further conversations are needed to evaluate the effectiveness and suggested outcomes of such a resource group, but we do encourage collaborations to set aside space for early career members to socialize, network, and collaborate with each other.


\subsection{Organizational Culture for Diversity, Equity and Inclusion}\label{sec:OrganizationalCulture}
    Institutions and collaborations should work to implement support structures, response practices, and other mechanisms to address the issues identified in this report. As identified in section \ref{sec:Institutions}, it is increasingly common for institutions to have one or more committees. Steps need to be taken to ensure that these are effective in meaningful changes including more careful organization and planning, structures of accountability, institutional power, and effective support from community/collaboration leadership. This should include organization of activities in the institution and high level involvement of management. Examples include the involvement of executive sponsors from the directorate in the committees, and leadership by a central authority such as a Chief Diversity Officer who has power in the organization. Evidence-based best practices should be sought and implemented by institutions in order to improve workforce diversity, recruiting, support, inclusion and equity and ensure that all have the best opportunity to excel once on board. They include:
    \begin{itemize}
        \item All search committee members required to take implicit bias training and discuss the lessons learned.
        \item Hiring committees required to have diverse membership, with e.g. requirements for female and under-represented group participation, and scientists from other Divisions are invited to serve in order to give their perspective.
        \item Job postings contain gender neutral language, have clear qualifications relevant to the position and do not inadvertently screen out under-represented minorities. 
        \item All positions are posted on diversity web sites to increase outreach to women and underrepresented minorities. 
    \end{itemize}
    
    Institutions should have active DEI committees, and employees in HEP areas should seek to serve in Employee Resources Groups (ERGs), such as the Women’s Science and Engineering Council, Early Career Resource Group, Lambda Alliance, Asian Pacific Islander, Veterans, African American and others. As part of increasing awareness and accountability of DEI principles, institutional leadership and staff should participate in a Equity Reset program with a equity curriculum addressing race, gender, gender and sexuality minorities and other aspects. We believe that one of the best ways to have diverse staff is to invest in the pipeline for talented and diverse students and encourage their careers. For example, staff in the HEP divisions actively participate in the SAGE workshops, a summer camp for high school girls that is co-hosted by SLAC and LBNL.\footnote{\url{https://k12education.lbl.gov/programs/high-school/sage}}  
    
    Implementing the various evidence based practices described in this section, has increased awareness and importance of DEI efforts in workforce development and resulted in actions. We employ numerous undergraduates who are in the pipeline toward STEM careers, and pride ourselves in the further development and career success of our diverse graduate students and postdocs. We need to continue these efforts and measure effectiveness of these practices. 
    
    The strategic goal: to create at all levels an institution where all employees feel valued, included, and empowered to reach their full potential.

%% file: CollaborationServices.tex
\recommendation{F1.4}{Funding and structural aid should be made available to develop ``Collaboration services'' offices at host laboratories that can provide HEPA collaborations and other physics communities with the following: a) advice on topics including those listed in F1.3, b) training in project management, ombudsperson roles, and victim-centered approaches to investigations, and c) logistical tools including facilitation of investigation and mediation.} 

\recommendation{C1.2}{The community should prioritize the implementation of best practices networks across institutions and communities of physics practice. This may be facilitated through Collaboration Services Offices (F1.4), but may also include the facilitation of networks between DEI groups at similar collaborations.}

\subsection{Development of network at host labs}
    
    Host laboratories can be assigned responsibility for services for their collaborations, examples of which include
    \begin{itemize}
        \item \textbf{Training.}  Host labs could provide DEI training to the collaborations they host.  Topics include microaggressions, conflict resolution, effective group management, accessibility (expecially in the post-COVID era). Labs already provide such training to their employees and employ experts in these fields. Initiatives such as LBNL Equity Reset\cite{--LBNLEquity-} have produced documentation for events that is already being propogated into other communities (namely the First Friday events have spread to the UC Berkeley Physics Department).  It is important at this point to note that not all collaboration members have access to training at their home institutions; for example, the DESI collaboration has invited LBNL diversity experts to speak at its meetings.
        \item \textbf{Reporting Inbox.}  Host labs can intake reports of concerns and violations in an anonymized way that is firewalled from the collaboration.  Fermilab provides this service for people with an account.

        \item \textbf{Personal conflict resolution.}  Collaborations generally do not have access to trained experts who can handle personal non-scientific conflicts. The availability of professional mediators would dramatically decrease the chance of additional harm compared to internal mediation.
        \item \textbf{Hosting Resources.} Myriad resources exist on a number of platforms by which individuals can educate themselves and help form more inclusive communities, so much so that for many people it may be overwhelming to know where to begin. For a large number of individuals in our community, rediscovering materials by word of mouth is likely not to be the most efficient and effective way of becoming educated. Furthermore, these materials may not be composed in ways that are conducive to application within particle physics at universities and national laboratories. Perhaps an individual is planning to hold a conference at their institution and wants to be inclusive but does not have ample time or resources to rediscover the most important aspects of inclusivity at a workshop of their variety. A solution should address questions and desires such as ``How should I educate myself or my group?” and ``I’m organizing a workshop, what am I missing?” By compiling go-to, literature-driven resources meant specifically for the physics community (``from physicists for physicists”) that are open source and living/evolving over the course of time, we can mitigate this problem to catalyze more widespread change and understanding of the issues.
    \end{itemize}
    
    Other services needed by collaborations may be better served at the community level, due to considerations of the economy of scale and independence. 
    \begin{itemize}
        \item \textbf{Investigation and adjudication.} Trusted impartial and independent bodies are sometimes needed to resolve incidents within collaborations . Professional societies and host laboratories should play a role.
        \item \textbf{Ombudsperson training} Training for collaboration ombudspeople is expensive, oversubscribed, and tends to be focused on an institutional rather than collaboration environment (which can differ dramatically, as demonstrated in sections \ref{sec:Institutions} and \ref{sec:Collaborations}. A network of community oriented offices can sponsor periodic collaboration-specific ombudsperson training workshops.
        \item \textbf{Climate surveys and external reviews. }The American Physical Society offers site visits for improving the climate for women and other marginalized groups in physics departments. Additionally, reviews can be prohibitively expensive for each collaboration, but some reviews are already ongoing at host labs (section \ref{sec:Implementation_nEXO}). Formal implementation, funded by host labs, can dramatically improve our access to data about the status and progress of climate related initiatives. 
    \end{itemize}
    
    \recommendation{F2.1}{Funding agencies should facilitate Climate Community Studies, instead of leaving such studies up to individual communities to complete. In line with F3.2, these studies should be informed by expertise in social and organizational dynamics.}
    
    \recommendation{F3.2}{In line with F2.1, community studies should receive advice from experts in sociology and organizational psychology, such that the tools used to evaluate the climate are adequate, effective, and informative. These studies and accompanying expertise should be funded at the federal and institutional levels. }

\subsection{Notes on implementation}

     Educating and training the community to bring awareness about the DEI issues is also essential for changing the culture. Again, this is something that is addressed to varying degrees by individual institutions but is not broadly accessible throughout the community. 
     
     Additionally, institutional training (specifically training on bias) done without care or the advice of experts has been shown to have little impact on either implicit or explicit bias \cite{forscherMetaanalysisProceduresChange2019, hagiwaraCallGroundingImplicit2020, pritloveGoodBadUgly2019}; therefore, as with so many things within the context of this contributed paper, such policies must be advised by DEI researchers who have expertise in interventions. Furthermore, training alone should not be represented as institutional change or commitment to DEI. Unfortunately, as discussed multiple places within this work, modern concepts and practices performed under the banner of `equity’, `diversity’, and `inclusion’ often fall far short of directly addressing injustice, and cannot be entrusted with driving the process for change\cite{prescodweinstein-Space+Anthropology-DiversityDangerous-2018}. 
     
     Instead, we encourage this collaboration services office to be driven by and/or staffed with experts in the fields of sociology and social/organizational psychology, and for policies made by these offices be in support of DEI beyond what is legally mandated by agency policy. 
    